\documentclass[journal]{IEEEtran}
\usepackage{epsfig}
\usepackage{amsmath,amssymb}
\usepackage{indentfirst}
\allowdisplaybreaks[4]
\usepackage{xcolor}
\usepackage{textcomp}
\usepackage{algorithm}
\usepackage{algorithmic}
\usepackage{multirow}
\usepackage{graphicx}
\usepackage{stfloats}
\usepackage{amsfonts}
\usepackage{mathrsfs}
\usepackage{tabu}
\usepackage[noblocks]{authblk}
\usepackage[compress]{cite}
\usepackage{bm}
\usepackage{longtable}
\usepackage{rotating}
\usepackage{chngpage}
\usepackage{array}
\usepackage{booktabs}
\usepackage[printonlyused]{acronym}
\usepackage{geometry}
\usepackage{amsthm,amsmath,amssymb}
\usepackage{mathrsfs}
\usepackage{enumerate}
\usepackage{amssymb}
\usepackage{verbatim}
\usepackage{subfigure}   
\usepackage{ulem}  
\usepackage{float} 
\usepackage{caption}
\DeclareMathOperator*{\argmax}{arg\,max}
\makeatletter

\newcommand{\Rmnum}[1]{\expandafter\@slowromancap\romannumeral #1@}
\makeatother
\begin{document}
\title{\huge Digital Twin Assisted Task Offloading for Aerial Edge Computing and Networks}
\author{Bin Li,~\IEEEmembership{Member,~IEEE}, Yufeng Liu, Ling Tan,~\IEEEmembership{Member,~IEEE}, Heng Pan, and Yan Zhang,~\IEEEmembership{Fellow,~IEEE}


\thanks{B. Li, Y. Liu, and L. Tan are with the School of Computer and Software, Nanjing University of Information Science and Technology, Nanjing 210044, China (e-mail: bin.li@nuist.edu.cn; yufengliu@nuist.edu.cn; cillatan0@nuist.edu.cn).}
\thanks{H. Pan is with Zhongyuan University of Technology, Zhengzhou 450007, China (e-mail: panheng@zut.edu.cn).}
\thanks{Y. Zhang is with University of Oslo, Norway, and also with Simula Metropolitan Center for Digital Engineering, Norway (e-mail: yanzhang@ieee.org).}
}

\markboth{IEEE Transactions on Vehicular Technology}
{Shell \MakeLowercase{\textit{et al.}}: Bare Demo of IEEEtran.cls for Journals}

\maketitle

\begin{abstract}
    Considering the user mobility and unpredictable mobile edge computing (MEC) environments, this paper studies the intelligent task offloading problem
    in unmanned aerial vehicle (UAV)-enabled MEC with the assistance of digital twin (DT).
	We aim at minimizing the energy consumption of the entire MEC system by jointly optimizing mobile terminal users (MTUs) association, UAV trajectory, transmission power distribution and computation capacity allocation while respecting the constraints of mission maximum processing delays.
    Specifically, double deep Q-network (DDQN) algorithm stemming from deep reinforcement learning is first proposed to effectively solve the problem of MTUs association and UAV trajectory.
    Then, the closed-form expression is employed to handle the problem of transmission power distribution and the computation capacity allocation problem is further addressed via an iterative algorithm. 
    Numerical results show that our proposed scheme is able to converge and significantly reduce the total energy consumption of the MEC system compared to the benchmark schemes. 
\end{abstract}


\section{Introduction \label{a}}
The proliferation of a variety of mobile services with rich experiences 
may bring unprecedented challenges to the computational performance of mobile devices due to their restricted calculation ability \cite{9262878,9163316,2020Novel}.
Although mobile edge computing (MEC) technology has been envisioned as a revolutionary solution to realize the ability of cloud computing at the edge of the networks \cite{9291470,9392259, 9485089}, there are still some deficiencies to be tackled, i.e., the location limitations of static ground base station (BS) and high deployment cost.
Owing to the flexible deployment and low price of unmanned aerial vehicles (UAVs) \cite{2020A,9420280}, UAVs as MEC nodes have emerged as a key advocate for providing mobile-edge services in $5$G emergency communications.

A number of research efforts have been dedicated to the UAV-enabled MEC for task offloading.
For instance, in \cite{9032145}, given the size of the calculation tasks and the deadline for completion, the authors considered the computation resources and UAV trajectory to minimize the weighted system energy consumption.
In \cite{8641189}, constrained by UAV's limited energy, the authors focused on applying UAV as an aerial BS to provide computational task offloading services to the ground users for maximizing migration throughput of user tasks.
The authors in \cite{9066989} investigated the total energy consumption of user equipments by jointly optimizing users’ association, uplink power control, and UAV 3-D placement, etc.
Nevertheless, the UAV-assisted MEC network also faces new challenges, i.e., the long line-of-sight (LoS) link \cite{9417457} between user and UAV will cause long transmission time and the computing resource at UAV is not always adequate, which are not friendly to the delay-sensitive tasks.

Benefiting from the proximity gain, device-to-device (D2D) communications have brought much attention in wireless research community. Recently, there has been growing interest in using a D2D communication link to offload the computing task to a nearby mobile device.
It is worth noting that some excellent work has been devoted to examine the performance of D2D offloading in MEC systems \cite{8643746,8751147,9029056}.  
These studies have demonstrated that D2D-aided computing offloading can not only be used to ensure low latency and high data transmission rate, but also to make up the problem of MEC server shortage. 
However, the mobility of users and the unpredictability of MEC environments in the real world have not yet been fully carried out. Thus, how to leveraging D2D offloading in UAV-enabled MEC while considering user mobility and 
time-varing environment is worthwhile to be investigated.

Providentially, digital twin (DT) as an emerging technology in the 6G era, which can digitize the real world, realizes the communication, cooperation and information sharing between the physical world and the virtual world, so as to create a mixed real virtual world. Moreover, DT creates virtual models to represent real objects in the physical network, monitors the status of the entire network in real time, and directly provides users with perceptual data to make more accurate and timely offloading decisions in a favorable way to meet the actual intelligent requirements \cite{9429703,2020Intelligent,9447819}.
In view of the advantages of DT, related work has combined DT and MEC to construct a Digital Twin Edge Network (DITEN). 
On one hand, DITEN collects data from various physical entities and stores it in devices dedicated to storing DT. On the other hand, it monitors the status of current network in real-time.

Compared with MEC network without DT, the offloading module does not always interact with the real-time environment and query the running status of each edge server, which not only improves the task offloading effectiveness and ensures user experience, but also reduces system energy consumption and saves system resources. However, the characteristics and uniqueness of DITEN make it face some challenges, such as scenarios for multiple mobile users and multiple computing platforms, how to place the DT, where to place it, and how to set the estimation errors.
The current research on DITEN is still in the early stage.
Typically, the authors of \cite{9170905} constructed the DITEN and proposed a permissioned blockchain empowered federated learning framework for realizing robust edge intelligence.
The authors in \cite{9174795} documented the issue of minimizing the latency in DITEN and treated BSs as the unload nodes for the mobile devices. 
Naturally, it is worth thinking about how to operate the system after the establishment of DITEN in detail  and ensure the quality of services (QoS) for all users once the number of computing tasks further increases.

Sparked by the above discussions, this paper considers an adaptive DITEN consisting of multiple mobile terminal users (MTUs), a UAV equipped with MEC server, multiple resource devices, and a BS, where the MTUs randomly generate computing tasks as they move. In this setting, our goal is to minimize the overall system energy consumption by jointly optimizing the MTUs’ association, UAV trajectory, transmission power distribution and the corresponding computation capacity allocation.

For the sake of clarity, the main contributions of this work are listed as follows:
\begin{itemize}
	\item 
	We propose an adaptive DT framework for a UAV-enabled MEC network to predict the network state accurately, where the user mobility is considered. We introduce the D2D communication links to help task offloading and develop an intelligent offloading strategy to manage the computing resource assignments. 
	\item 
	We study the joint optimization problem of MTUs association, UAV trajectory, transmission power distribution and computation capacity allocation from the view of energy consumption, subject to the constraints of delay.
	The formulated problem is a mixed integer nonlinear optimization problem and is very challenging to solve. To this end, we present transformations to reformulate the original problem into a tractable and achieve a near-optimal solution with remarkably reduced complexity.
	\item 
	We present a deep reinforcement learning (DRL) approach to find the MEC offloading decisions, and develop an iterative algorithm to optimize the computation
	capacity variable.
	Meanwhile, the convergence property and the computational complexity are examined.
	Extensive numerical results evaluate the effectiveness and superiority of our proposed method in reducing system energy consumption. 
	Moreover, our work provides a low-complexity design guideline for 
	MTUs to complete tasks. 
\end{itemize}
\begin{figure}[t]
	\includegraphics[width=3.45in]{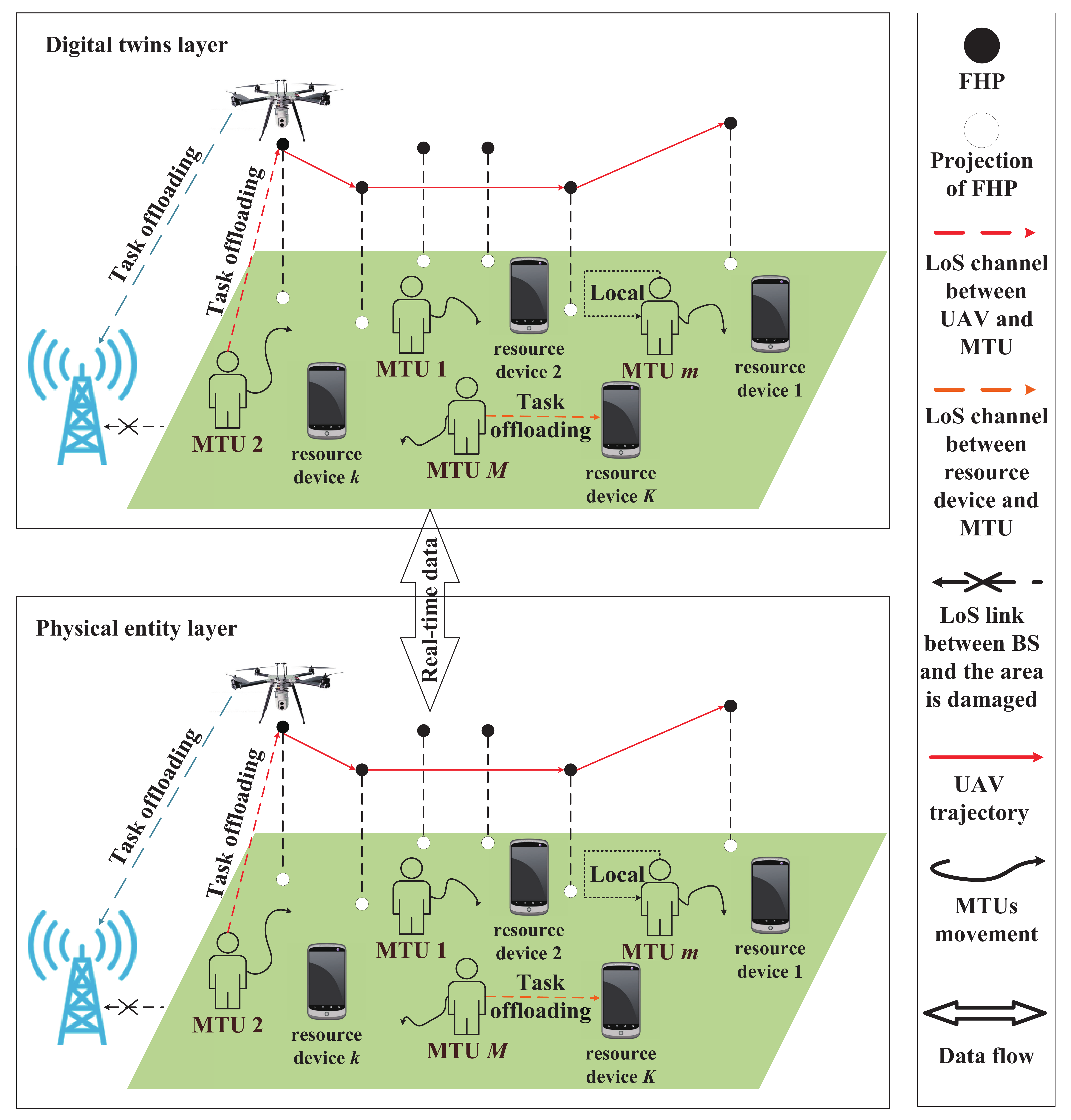}
	\caption{Intelligent offloading architecture of UAV-assisted DITEN.}
	\label{Model}
\end{figure}
The remainder of this article is organized as follows. 
The related work is discussed in Section II.
In Section III, we describe the system model and formulate the optimization problem. In Section IV, we present the solution to the considered problem. In Section V, we analyze the computational complexity and the convergence of the proposed algorithm. The numerical results and conclusions are given in detail in Section VI and VII, respectively.

\section{Related Work \label{h}}
In this section, we introduce the related work on considering the integration of MEC and DT to build a DITEN.

As an illustration, in \cite{9244624}, the authors proposed the application of DT technology to the industrial internet of things, in which DT can map the real-time operating status and behavior of equipments to the digital world, and finally realize the dynamic tradeoff between computing energy consumption and communication energy consumption in time-varying communication environment.
In \cite{9451579}, the authors combined DT technology with artificial intelligence effectively, and applied it to the design of automotive edge computing network. The DT technology can help reveal the potential edge service matching between large-scale vehicle pairs and effectively reduce the complexity of service management. 
In addition, the authors in \cite{9145588} elaborated the combination of DT network and mobile edge network from different perspectives, as well as further introduced the asynchronous model update scheme. This work improves the communication efficiency of the system and reduces the total energy consumption.
At the same time, the work in \cite{9447819} described a DT-assisted task offloading scheme, including the selection of mobile edge server and task offloading, and finally realized the two-way reduction of energy consumption and time cost. Additionally, in \cite{9399641}, the authors utilized DT to describe the vehicular social relations and constructed a social model. Furthermore, a deep learning-based scheme was proposed to optimize caching decisions for the maximization of system utility.
In \cite{9786719}, the authors proposed a DT framework and jointly considered the communication, computing, and storage, which effectively achieves the goal of minimizing the latency performance.
These studies provide useful insights in DITEN, but they mainly focus on the ground DITEN.

Compared to the above mentioned discussions, UAV-assisted DITEN has also been studied in existing literature. 
As a pioneering work, the authors in \cite{9128981} established a DT model for UAV-assisted MEC system, in which a UAV is implemented as a complementary computing server flying in the air and the long-term computation performance is maximized.
The work in \cite{9311405} considered dynamic DT and federated learning for air-ground networks with the aim of improving accuracy and energy efficiency. 
In order to capture the time-varying resource status of physical objects,
the authors of \cite{9351542} designed an aerial-assisted internet of vehicles supported by DT, where a two-stage incentive mechanism for resource
allocation is considered to reduce delay and improve energy efficiency. 

The aforementioned studies have proved that the combined research of DT and MEC is extremely valuable. 
Nevertheless, to our knowledge, there has been little research
on air-ground collaboration in DITEN. In this paper, we consider the problem of task offloading in a UAV-enabled MEC with user mobility and apply the framework of adaptive DT to predict the dynamic network environment accurately.

\section{System Model and Problem Formulation \label{b}}

We consider a UAV-assisted DITEN, as shown in Fig. \ref{Model}.
In the physical entity layer, there are $M$ MTUs, $K$ resource devices, a UAV and one base station (BS).
We define $\mathcal{M} = \left\{1,..., M\right\}$ as the set of MTUs and $\mathcal{K} = \left\{1,..., K\right\}$ as the set of resource devices, respectively.
Assume that the movement cycle of all MTUs is $T$ and they are randomly divided in a region when $T=0$.
In the process of moving, each of MTUs may generate some deadline-sensitive computation tasks to be completed, which are independent and cannot be divided.
Considering that some MTUs are limited by computing capacity or battery life and cannot effectively complete the caculation tasks, 
we propose that neighbor resource devices, UAV and BS are used to assist MTUs to finish tasks.
Thereinto, the UAV can only hover one of $Q$ fixed hover points (FHPs) in each time slot, meanwhile, the LoS link between BS and MTU's active area is blocked. If MTU $m$ wants to offload the mission to the BS, the UAV acts as a relay to help the task offloading.
Consequently, the tasks on each MTU can be completed in one of the following four ways, i.e., local computing, offloading to resource devices, unloading to UAV or BS.
In the digital twins layer, the digital representation of physical entities,
e.g., MTUs, resource devices, UAV, provides the dynamics of how the system operates.

\begin{figure}[t]
	\includegraphics[width=3.4in]{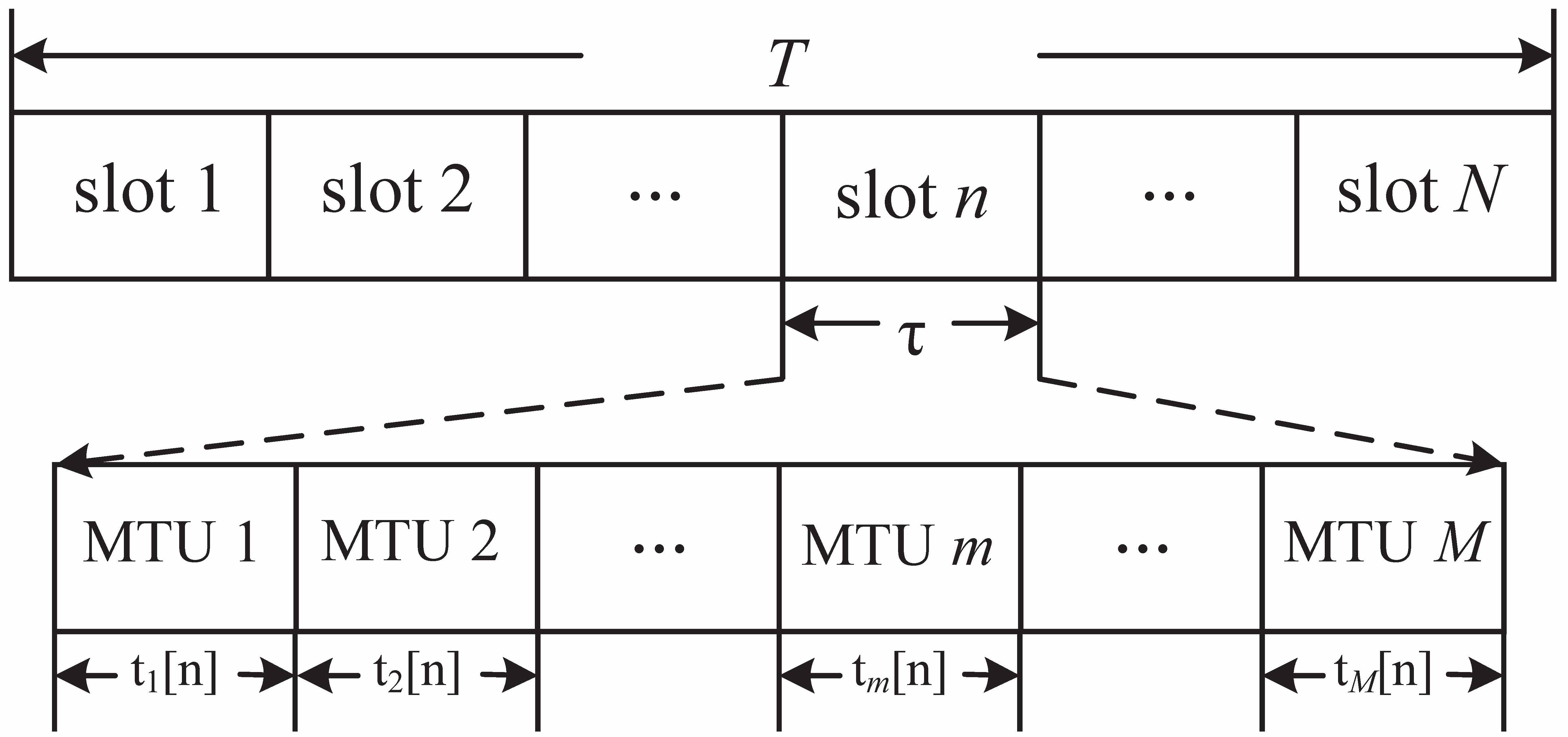}
	\caption{The time slot division protocol for MTUs in TDMA scheme.}
	\label{Slot}
\end{figure}

In order to avoid interference and accurately capture the locations of all MTUs, the time-division multiple access (TDMA) \cite{9203867} protocol is employed for multiple MTUs offloading computation tasks as soon as possible to their nearby resource devices, UAV, and BS.
In TDMA-based task offloading, we divide MTUs' mobility trajectory over time $T$ in $N$ time slots of equal duration $\tau$ such that $T = N \tau$, as shown in Fig. \ref{Slot}. 
For notational convenience, the set of time slots is represented as $\mathcal{N} = \left\{1,..., N\right\}$. Also, each time slot $n \in \mathcal{N}$ is further divided into $M$ durations that the operations related to MTU $m$ are all executed in the $m$-th duration $t_m[n] \in \left[0,\tau\right]$, satisfying the following constraint
{
	\setlength{\abovedisplayskip}{0.2cm}
	\setlength{\belowdisplayskip}{0.2cm}
	\begin{align}
	\sum_{m=1}^M t_{m}[n] \leq \tau,~\forall n\in\mathcal{N}.
	\label{one}
	\end{align} 
}\indent For a three dimensional Cartesian coordinate system, without loss of generality, we consider that the location of resource device $k$ is denoted as $\mathbf{l}_{k}=[x_{k},y_{k},0]^{T},~\forall k \in \mathcal{K}$,
the UAV flies at a constant height $H$ ($H$ \textgreater $~0$) and its horizontal position in time slot $n\in \mathcal{N}$ can be expressed as $\mathbf{l}_{j}^{\rm UAV}[n]=[x_{j}^{\rm UAV}[n],y_{j}^{\rm UAV}[n],H]^{T}, ~\forall j\in \mathcal{J'}, \mathcal{J'} = \left\{K+1,..., K+Q\right\}$, and the corresponding horizontal of BS is written as $\mathbf{l}_{0}^{\rm BS}=[x_{0}^{\rm BS},y_{0}^{\rm BS},0]^{T}$.
In time slot $n$, the computation tasks of MTU $m$ can be described as
{
\setlength{\abovedisplayskip}{0.2cm}
\setlength{\belowdisplayskip}{0.2cm}
\begin{align}
U_m[n] = \left\{D_m[n],C_m[n],T_m[n] \right\},~\forall m \in \mathcal{M}, ~\forall n \in \mathcal{N},
\end{align}
}\noindent where $D_m[n]$ represents the number of bits for the computation task, $C_m[n]$ indicates the number of CPU cycles required to complete the computation task of $1$-bit, and $T_m[n]$ is the maximum allowable latency to complete the task.
\vspace{-5pt}
\subsection{A DT Model}
In this paper, we consider three categories of DTs, such as twin MTUs, twin resource devices and twin UAV. 
These DTs are established in the BS to store the raw data of each network entity and monitor the current operating state of the network.

The DT of an MTU is a digital replica of the MTU, which interacts constantly with the MTU and updates itself with the actual network topology, the requests of tasks, etc.
Note that DTs can’t fully reflect the MTUs’ state and may have an estimated deviation from the true value of the MTU state.
In this framework, we use the deviation of available CPU frequency
$\widetilde{f}_{m}^{l}[n]$ to describe the deviation between real MTU $m$ and its DT in time slot $n$, which can be either positive or negative. For MTU $m$, its digital twin $D_{m}^{\rm MTU}[n]$ in the $n$-th time slot can be expressed as
{
	\setlength{\abovedisplayskip}{0.2cm}
	\setlength{\belowdisplayskip}{0.2cm}
	\begin{align}
	D_{m}^{\rm MTU}[n] = \Theta(f_{m}^{l}[n],\widetilde{f}_{m}^{l}[n]),~\forall m \in \mathcal{M},~\forall n \in \mathcal{N},
	\end{align}
}\noindent where $f_{m}^{l}[n]$ is the estimated computation capability (in CPU cycles per second) of MTU $m$ stored in DT.

For ease of exposition, we define $f_{j,m}[n]$ as the estimated CPU cycle frequency of resource device $j$ and its digital twin $D_{j}[n]$ can be given by
{
	\setlength{\abovedisplayskip}{0.2cm}
	\setlength{\belowdisplayskip}{0.2cm}
	\begin{align}
	D_{j}[n] = \Theta(f_{j,m}[n],\widetilde{f}_{j,m}[n]),~\forall j \in \mathcal{K},~\forall m \in \mathcal{M},
	\end{align}
}\noindent where $\widetilde{f}_{j,m}[n]$ is the deviation of available CPU frequency between resource device $j$ and its DT.

Similarly, the deviation of available CPU frequency between UAV and its DT is denoted by $\widetilde{f}_{j,m}^{\rm UAV}[n]$. As a consequence, the digital twin $D_{j}^{\rm UAV}[n]$ can be described by
{
	\setlength{\abovedisplayskip}{0.2cm}
	\setlength{\belowdisplayskip}{0.2cm}
	\begin{align}
	D_{j}^{\rm UAV}[n] = \Theta(f_{j,m}^{\rm UAV}[n],\widetilde{f}_{j,m}^{\rm UAV}[n]),~\forall j \in \mathcal{J'},~\forall m \in \mathcal{M},
	\end{align}
}\noindent where $f_{j,m}^{\rm UAV}[n]$ is the estimated CPU cycle frequency of UAV $j$ stored in DT.
\vspace{-9pt}
\subsection{Mobility Model of MTUs}
We consider that all MTUs are randomly located in time slot $n = 0$ and all MTUs' locations do not change during the duration $\Delta_{n,n-1}$ between the $(n-1)$-th and $n$-th time slot. According to the Gauss-Markov random model (GMRM) described in \cite{9044434}, the velocity $v_m[n]$ and direction $\theta_m[n]$ of the $m$-th MTU in each time slot $n$ $(n \geq 1)$ are updated as
{
	\setlength{\abovedisplayskip}{0.2cm}
	\setlength{\belowdisplayskip}{0.2cm}
	\begin{align}  
	v_m[n] = \mu_1 v_m[n-1] + (1- \mu_1) \overline{v} + \sqrt{1-{\mu_1}^2} \Lambda_m, 
	\label{two}  \\
	\theta_m[n] = \mu_2 v_m[n-1] + (1- \mu_2) \overline{\theta_m} + \sqrt{1-{\mu_2}^2} \Gamma_m, 
	\label{three}
	\end{align} 
}where $0 \leq \mu_1,\mu_2  \leq 1$ are used to adjust for the effects of the previous state, $\overline{v}$ represents the average speed of all MTUs, and  $\overline{\theta_m}$ is the average direction of the $m$-th MTU. 
In particular, we consider that the average speed for all MTUs is same and different MTUs have different average directions. 
Also, $\Lambda_m$ and $\Gamma_m$ follow two independent Gaussian distributions with different mean-variance pairs $({\overline{\xi}}_{v_m},{\overline{\zeta}}_{v_m}^2)$ and
$({\overline{\xi}}_{\theta_m},{\overline{\zeta}}_{\theta_m}^2)$ for the $m$-th MTU, both of which reflect the randomness in the movements of different MTUs.
In addition, we use $\mathbf l_{m}^{\rm MTU}[n]= [x_{m}^{\rm MTU}[n],y_{m}^{\rm MTU}[n],0]^{T}$ to represent the location of the $m$-th MTU in the $n$-th time slot.
On the basis of $(\ref{two})$ and $(\ref{three})$, the location of MTU $m$ is updated as follows
{
	\setlength{\abovedisplayskip}{0.2cm}
		\begin{align}
		x_{m}^{\rm MTU}[n] = x_{m}^{\rm MTU}[n-1] + v_m[n-1]cos(\theta_m[n-1])t_m[n], 
		\label{four}  \\
		y_{m}^{\rm MTU}[n] = y_{m}^{\rm MTU}[n-1] + v_m[n-1]sin(\theta_m[n-1])t_m[n].
		\label{five}
		\end{align} 
}
\vspace{-30pt}
\subsection{Task Computation Model}
In each time slot $n$, the set of possible servers that help MTUs completing the calculation is denoted by $\mathcal{J}=\left\{-1,0,1,2,...,K,K+1,K+2,...,K+Q\right\}$. We define a decision variable $\alpha_{m}^{j}[n] \in \left\{0,1\right\}$, $~\forall j \in \mathcal{J}$ to distinguish different offloading patterns, where the subscript $m$ represents MTU $m$, and the superscript $j$ indicates the $j$-th server. 
If the task is executed by itself, $\alpha_{m}^{-1}[n] = 1$. Similarly, $\alpha_{m}^{0}[n] = 1$ represents that the task of MTU $m$ is offloaded to BS by UAV,  $\alpha_{m}^{j}[n] = 1$ for $~\forall j \in \mathcal{K}$ indicates that the task of MTU $m$ is offloaded to the $j$-th resource device, and $\alpha_{m}^{j}[n] = 1$ for $~\forall j \in \mathcal{J^{'}}$ represents that the task of MTU $m$ is offloaded to UAV hovering over $j$.
Since each task is independent and can only be performed in one place, we have
{
	\setlength{\abovedisplayskip}{0.2cm}
	\setlength{\belowdisplayskip}{0.2cm}
	\begin{align}
		\sum_{j=-1}^{Q+K} a_{m}^{j}[n] = 1,~\forall m\in\mathcal{M}, ~\forall n\in\mathcal{N}.
	\end{align}
}\indent Based on MTUs' offloading decisions, the computation latency of a task is determined by the local execution or offloading delays, 
where the latter involves transmission as well as remote execution time. We will discuss the model for each process in the following sections.

\subsubsection{\textbf{Local Computing}}
If the offloading cost is high, local execution is preferred. 
The estimated time required to perform the task of MTU $m$ in time slot $n$ is expressed as {\cite{9691475, 3-2}
{
	\setlength{\abovedisplayskip}{0.2cm}
	\setlength{\belowdisplayskip}{0.2cm}
	\begin{align}
	\widetilde{T}_{m}^{l}[n]=(D_{m}[n]C_{m}[n]) / f_{m}^{l}[n].
	\label{six} 
	\end{align}
}\indent Note that DT can’t fully reflect the MTU’s state and may have an estimated deviation. Assuming that the deviation between MTU $m$ and its DT can be acquired in advance, the computing latency gap $\Delta T_m^{l}[n]$ between real value and DT estimation can be calculated by {\cite{9691475, 3-2}
{
	\setlength{\abovedisplayskip}{0.2cm}
	\setlength{\belowdisplayskip}{0.2cm}
	\begin{align}
		\Delta T_{m}^{l}[n] = \frac{D_{m}[n]C_{m}[n]\widetilde{f}_{m}^{l}[n]}{f_{m}^{l}[n]({f_{m}^{l}[n]-\widetilde{f}_{m}^{l}[n]})}.
		\label{seven}
	\end{align}
}\indent Then, the computing latency of the $m$-th MTU for local execution can be represented by
{
	\setlength{\abovedisplayskip}{0.2cm}
	\setlength{\belowdisplayskip}{0.2cm}
	\begin{align}
		T_{m}^{l}[n] = \widetilde{T}_{m}^{l}[n] + \Delta T_{m}^{l}[n].
		\label{eight}
	\end{align}
}\indent Furthermore, the local caculating energy consumption of MTU $m$ is given by
{
	\setlength{\abovedisplayskip}{0.2cm}
	\setlength{\belowdisplayskip}{0.2cm}
	\begin{align}
	E_{m}^{l}[n] = k_m(f_{m}^{l}[n]-\widetilde{f}_{m}^{l}[n])^{2}C_{m}[n]D_{m}[n],
	\label{nine} 
	\end{align} 
}\noindent where $k_m$ is the effective switching capacitance coefficient depending on the hardware performance of MTU $m$ \cite{9435770}.

\subsubsection{\textbf{Offloading to Resource Devices for Computing}}
Once $a_{m}^{j}[n]=1,~\forall j \in \mathcal{K}$, i.e., the task of MTU $m$ is offloaded to resource device $j$ in time slot $n$ and executed remotely. 
Generally, the computation results are relatively much smaller in size and the results downloading time can be ignored \cite{9481222}.
In terms of the coordinates, the distance between MTU $m$ and resource device $j$ in time slot $n$ can be represented as
{
	\setlength{\abovedisplayskip}{0.2cm}
	\setlength{\belowdisplayskip}{0.2cm}
	\begin{align}
	d_{m,j}[n]=\|\mathbf{l}_{m}^{\rm MTU}[n]-\mathbf{l}_{j}\|,~\forall m\in\mathcal{M},~\forall j\in\mathcal{K}.
	\label{ten} 
	\end{align} 
}\indent Similar to \cite{9333960}, in the $n$-th time slot, the channel power gain from MTU $m$ to resource device $j$  can be expressed as
{
	\setlength{\abovedisplayskip}{0.2cm}
	\setlength{\belowdisplayskip}{0.2cm}
	\begin{align}
		h_{m,j}[n]= \beta_{0} (d_{m,j}[n])^{-2},~\forall m\in\mathcal{M},~\forall j\in\mathcal{K},
		\label{ten-1}
	\end{align}
}\noindent where $\beta_{\mathrm{0}}$ represents the channel power gain at a reference distance of one meter \cite{9566305}.

Hence, the achievable transmission rate from MTU $m$ to resource device $j$ in time frame $n$ can be calculated as 
{
	\setlength{\abovedisplayskip}{0.2cm}
	\setlength{\belowdisplayskip}{0.2cm}
	\begin{align}
	\nonumber R_{m,j}[n] = B\log_2(1 + \frac{p_{m,j}[n] h_{m,j}[n]}{\sigma^2}),
	~\forall m\in\mathcal{M},\\~\forall j\in\mathcal{K},~\forall n\in\mathcal{N},
	\label{ten-2} 
	\end{align} 
}\noindent where $B$ is the bandwidth of the system, $p_{m,j}[n]$ denotes the transmit power of MTU $m$ to resource device $j$ for offloading tasks in time slot $n$, and
$\sigma^2$ signifies the additive white Gaussian noise power \cite{9184017}.
According to (\ref{ten-2}), the duration of transmission tasks of MTU $m$ is given by
{
	\setlength{\abovedisplayskip}{0.2cm}
	\setlength{\belowdisplayskip}{0.2cm}
	\begin{align}
	T_{m,j}[n]= \frac{D_{m}[n]}{R_{m,j}[n]},
	~\forall m\in\mathcal{M},~\forall j\in\mathcal{K}.
	\label{ten-3} 
	\end{align}
}\indent Correspondingly, the energy consumption due to transmission can be obtained as
{
	\setlength{\abovedisplayskip}{0.2cm}
	\setlength{\belowdisplayskip}{0.2cm}
	\begin{align}
	E_{m,j}[n] = p_{m,j}[n] T_{m,j}[n],
	~\forall m\in\mathcal{M},&~\forall j\in\mathcal{K}.
	\label{E-RD} 
	\end{align}
}\indent In time slot $n$, the estimated processing delay incurred by performing task $D_{m}[n]$ is defined as
{
	\setlength{\abovedisplayskip}{0.2cm}
	\setlength{\belowdisplayskip}{0.2cm}
	\begin{align}
	\widetilde{T}_{j,m}[n]=(D_{m}[n]C_{m}[n]) / f_{j,m}[n], ~\forall m\in\mathcal{M}, ~\forall j\in\mathcal{K}.
	\label{ten-5} 
	\end{align}
}\indent Meanwhile, the computing latency gap $\Delta T_{j,m}[n]$ between real value and DT estimation is given by
{
	\setlength{\abovedisplayskip}{0.2cm}
	\setlength{\belowdisplayskip}{0.2cm}
	\begin{align}
	\nonumber \Delta T_{j,m}[n] = \frac{D_{m}[n]C_{m}[n]\widetilde{f}_{j,m}[n]}{f_{j,m}[n]({f_{j,m}[n]-\widetilde{f}_{j,m}[n]})},
	~\forall m\in\mathcal{M},\\~\forall j\in\mathcal{K},~\forall n\in\mathcal{N}.
	\label{ten-6}
	\end{align} 
}\indent Then, the actual computation latency at resource device $j$ side to process the offloaded tasks from MTU $m$ can be calculated as
{
	\setlength{\abovedisplayskip}{0.2cm}
	\setlength{\belowdisplayskip}{0.2cm}
	\begin{align}
	T_{j,m}[n] = \widetilde{T}_{j,m}[n] + \Delta T_{j,m}[n],~\forall m\in\mathcal{M},~\forall j\in\mathcal{K}.
	\label{ten-7}
	\end{align}
}\indent After receiving the offloaded task data from MTU $m$, resource device $j$ can start the computation process. Accordingly, the energy consumed by resource device $j$ to calculate the task of MTU $m$ can be expressed as
{
	\setlength{\abovedisplayskip}{0.2cm}
	\setlength{\belowdisplayskip}{0.2cm}
	\begin{align}
	E_{j,m}[n] = k_{j}(f_{j,m}[n]-\widetilde{f}_{j,m}[n])^{2}C_{m}[n]D_{m}[n],
	~\forall j\in\mathcal{K},
	\label{ten-8} 
	\end{align}
}\noindent where $k_{j}$ represents the effective switching capacitance coefficient of resource device $j$ \cite{9435770}.

\subsubsection{\textbf{Offloading to UAV for Computing}}
If $a_{m}^{j}[n]=1, ~\forall j \in \mathcal{J'}$, MTU $m$ will unload the computing task to UAV hovering over $j$. As a consequence, the distance between MTU $m$ and UAV hovering over $j$ can be represented as
{
	\setlength{\abovedisplayskip}{0.2cm}
	\setlength{\belowdisplayskip}{0.2cm}
	\begin{align}
	d_{m,j}^{\rm UAV}[n]=\|\mathbf{l}_{m}^{\rm MTU}[n]-\mathbf{l}_{j}^{\rm UAV}[n]\|,~\forall m\in\mathcal{M},~\forall j\in\mathcal{J^{'}}.
	\label{ten-9} 
	\end{align} 
}\indent As such, the LoS channel power gain from MTU $m$ to UAV hovering over $j$ in the $n$-th time slot follows the free space path loss model, given by
{
	\setlength{\abovedisplayskip}{0.2cm}
	\setlength{\belowdisplayskip}{0.2cm}
	\begin{align}
	h_{m,j}^{\rm UAV}[n]= \beta_{0} (d_{m,j}^{\rm UAV}[n])^{-2}, \forall m\in\mathcal{M},~\forall j\in\mathcal{J^{'}}.
	\label{twenty}
	\end{align}
}\indent Accordingly, the transmission rate from MTU $m$ to UAV hovering over $j$, denoted by $R_{m,j}^{\rm UAV}[n]$, is computed as follows
{
	\setlength{\abovedisplayskip}{0.2cm}
	\begin{align}
	\nonumber R_{m,j}^{\rm UAV}[n] = B\log_2(1 + \frac{p_{m,j}^{\rm UAV}[n] h_{m,j}^{\rm UAV}[n]}{\sigma^2}),
	\\~\forall m\in\mathcal{M},~\forall j\in\mathcal{J^{'}},~\forall n\in\mathcal{N},
	\label{twenty-1} 
	\end{align}
}\noindent where $p_{m,j}^{\rm UAV}[n]$ denotes the transmit power from MTU $m$ to UAV hovering over $j$ for offloading in time slot $n$. 
 
Then, the time required to send task of MTU $m$ is given by
{
	\setlength{\abovedisplayskip}{0.2cm}
	\begin{align}
	T_{m,j}^{\rm{UAV}-trans}[n]= \frac{D_{m}[n]}{R_{m,j}^{\rm UAV}[n]},
	~\forall m\in\mathcal{M},~\forall j\in\mathcal{J^{'}},~\forall n\in\mathcal{N},
	\label{twenty-2} 
	\end{align} 
}\noindent and the corresponding energy consumption of MTU $m$ due to offload task is defined as
{
	\setlength{\abovedisplayskip}{0.2cm}
	\begin{align}
	E_{m,j}^{\rm UAV}[n] = p_{m,j}^{\rm UAV}[n] T_{m,j}^{\rm UAV-trans}[n],
	~\forall m\in\mathcal{M},~\forall j\in\mathcal{J^{'}}.
	\label{twenty-3} 
	\end{align} 
}\indent The estimated time required to finish task $U_{m}[n]$ can be expressed as
{
	\setlength{\abovedisplayskip}{0.2cm}
	\setlength{\belowdisplayskip}{0.2cm}
	\begin{align}
	\nonumber \widetilde{T}_{j,m}^{\rm UAV}[n]=(D_{m}[n]C_{m}[n]) / f_{j,m}^{\rm UAV}[n],~\forall m\in\mathcal{M},\\~\forall j\in\mathcal{J^{'}},~\forall n\in\mathcal{N}.
	\label{twenty-4} 
	\end{align}
}\indent Correspondingly, the deviation between UAV and its DT is given ahead of time, so the computing time gap $\Delta T_{j,m}^{\rm UAV}[n]$
between real value and DT estimation can be obtained as
{
	\setlength{\abovedisplayskip}{0.2cm}
	\begin{align}
	\nonumber \Delta T_{j,m}^{\rm UAV}[n] = \frac{D_{m}[n]C_{m}[n]\widetilde{f}_{j,m}^{\rm UAV}[n]}{f_{j,m}^{\rm UAV}[n]({f_{j,m}^{\rm UAV}[n]-\widetilde{f}_{j,m}^{\rm UAV}[n]})},
	\\~\forall m\in\mathcal{M},~\forall j\in\mathcal{J^{'}}.
	\label{twenty-5}
	\end{align}
}\indent Hence, the actual computing time for processing task from MTU $m$ at the UAV can be given by
{
	\setlength{\abovedisplayskip}{0.2cm}
	\setlength{\belowdisplayskip}{0.2cm}
	\begin{align}
	\nonumber T_{m}^{\rm UAV}[n] = \widetilde{T}_{j,m}^{\rm UAV}[n] + \Delta T_{j,m}^{\rm UAV}[n],\\~\forall m\in\mathcal{M},~\forall j\in\mathcal{J^{'}}.
	\label{twenty-6}
	\end{align} 
}\indent Additionally, the energy incurred by the UAV is computed as
{
	\setlength{\abovedisplayskip}{0.2cm}
	\setlength{\belowdisplayskip}{0.2cm}
	\begin{align}
	E_{j,m}^{\rm UAV}[n] = k_Q(f_{j,m}^{\rm UAV}[n]-\widetilde{f}_{j,m}^{\rm UAV}[n])^{2}C_{m}[n]D_{m}[n],
	\label{twenty-7}
	\end{align} 
}\noindent where $k_Q = 10 ^{-26}$ stands for the factor about switched capacitance
of UAV \cite{9435770}.

\begin{itemize}
	\item
	\textbf{UAV Flying Energy Consumption:}
    UAV flying energy consumption $E_{fly}^{\rm UAV}[n]$ is only determined by the flight distance from one FHP to the other one during time slot $n$, then it can be denoted as
    {
    	\setlength{\abovedisplayskip}{0.2cm}
    	\setlength{\belowdisplayskip}{0.2cm}
    	\begin{align}
    	\nonumber E_{fly}^{\rm UAV}[n] = &P_f
    	((x_{j}^{\rm UAV}[n]-x_{j}^{\rm UAV}[n-1])^2+\\&(y_{j}^{\rm UAV}[n]-y_{j}^{\rm UAV}[n-1])^2)^{1/2}V^{-1},
    	\label{twenty-8} 
    	\end{align}
    }\noindent where $P_f$ and  $V$ is the flight power and flight speed, respectively.
	\item
	\textbf{UAV Hovering Energy Consumption:}
	When UAV flies to the $j$-th point, it should hover there until completing the offloading task $U_{m}[n]$. Consequently, UAV hovering energy consumption can be given as
	{
		\setlength{\abovedisplayskip}{0.2cm}
		\setlength{\belowdisplayskip}{0.2cm}
		\begin{align}
		E_{hov}^{\rm UAV}[n] = P_h (T_{m,j}^{\rm UAV-trans}[n]+T_{m,j}^{\rm UAV}[n]),
		\label{twenty-9} 
		\end{align}
	}\noindent where $P_h$ represents the UAV hovering power.
\end{itemize}

\subsubsection{\textbf{Offloading to BS for Computing}}	
Due to limited resources, UAV may further offload MTUs’ tasks to the more powerful BS. Based on the location of MTU $m$ in time slot $n$, the distance between MTU $m$ and UAV hovering over $j$ can be represented as
{
	\setlength{\abovedisplayskip}{0.2cm}
	\setlength{\belowdisplayskip}{0.2cm}
	\begin{align}
	d_{m-j}[n]=\|\mathbf{l}_{m}^{\rm MTU}[n]-\mathbf{l}_{j}^{\rm UAV}[n]\|,~\forall m\in\mathcal{M},~\forall j\in\mathcal{J^{'}}.
	\label{thirty} 
	\end{align}
}\indent The channel power gain from MTU $m$ and UAV hovering over $j$ is described by
{
	\setlength{\abovedisplayskip}{0.2cm}
	\setlength{\belowdisplayskip}{0.2cm}
	\begin{align}
	h_{m-j}[n]= \beta_{0} (d_{m-j}[n])^{-2},~\forall m\in\mathcal{M},~\forall j\in\mathcal{J^{'}}.
	\label{thirty-1}
	\end{align}
}\indent According to Shannon formula, we can compute the transmission rate for offloading task as
{
	\setlength{\abovedisplayskip}{0.2cm}
	\setlength{\belowdisplayskip}{0.2cm}
	\begin{align}
	\nonumber R_{m-j}^{\rm trans}[n] = B\log_2(1 +& \frac{p_{m,j}^{\rm UAV}[n] h_{m-j}[n]}{\sigma^2}),
	~\forall m\in\mathcal{M},\\
	&~\forall j\in\mathcal{J^{'}},~\forall n\in\mathcal{N}.
	\label{thirty-2} 
	\end{align} 
}\indent Then, the time required to offload task of MTU $m$ is given by
{
	\setlength{\abovedisplayskip}{0.2cm}
	\setlength{\belowdisplayskip}{0.2cm}
	\begin{align}
	T_{m-j}^{\rm trans}[n]= \frac{D_{m}[n]}{R_{m-j}^{\rm trans}[n]},
	~\forall m\in\mathcal{M},~\forall j\in\mathcal{J^{'}},~\forall n\in\mathcal{N},
	\label{thirty-3} 
	\end{align} 	
}\noindent and the energy consumed by MTU $m$ for offloading task can be modeled as
{
	\setlength{\abovedisplayskip}{0.2cm}
	\setlength{\belowdisplayskip}{0.2cm}
	\begin{align}
	E_{m-j}^{\rm trans}[n] = p_{m,j}^{\rm UAV}[n] 	T_{m-j}^{\rm trans}[n],
	~\forall m\in\mathcal{M},~\forall j\in\mathcal{J^{'}}.
	\label{thirty-4} 
	\end{align} 
}\indent Similarly, the transmission rate between UAV hovering over $j$ and BS in time slot $n$ can be calculated as 
{
	\setlength{\abovedisplayskip}{0.2cm}
	\setlength{\belowdisplayskip}{0.2cm}
	\begin{align}
	\nonumber R_{j-B}^{\rm trans}[n] = B\log_2(1 + \frac{p_{j,B}^{\rm BS}[n] h_{j-B}[n]}{\sigma^2}),\\
	~\forall j\in\mathcal{J^{'}},~\forall n\in\mathcal{N},
	\label{thirty-9} 
	\end{align} 
}\noindent where $p_{j,B}^{\rm BS}[n]$ indicates the transmit power of UAV hovering over $j$ to BS and $h_{j-B}[n]$ is the channel power gain from UAV hovering over $j$ to BS.
\\ \indent Moreover, the time required to unload the task from the UAV hovering at $j$ to BS is given by
{
	\setlength{\abovedisplayskip}{0.2cm}
	\setlength{\belowdisplayskip}{0.2cm}
	\begin{align}
	T_{j-B}^{\rm trans}[n]= \frac{D_{m}[n]}{R_{j-B}^{\rm trans}[n]},
	~\forall j\in\mathcal{J^{'}},~\forall n\in\mathcal{N},
	\label{forty} 
	\end{align} 
}\noindent and the energy consumption consumed by the UAV hovering at $j$ to transmit the task to BS can be expressed as
{
	\setlength{\abovedisplayskip}{0.2cm}
	\setlength{\belowdisplayskip}{0.2cm}
	\begin{align}
	E_{j-B}^{\rm trans}[n] = p_{j,B}^{\rm BS}[n] T_{j-B}^{\rm trans}[n],
	&~\forall m\in\mathcal{M},~\forall j\in\mathcal{J^{'}}.
	\label{forty-1} 
	\end{align} 
}\indent In the whole process, UAV flying energy consumption $E_{fly1}^{\rm UAV}[n]$ can be denoted by 
{
	\setlength{\abovedisplayskip}{0.2cm}
	\setlength{\belowdisplayskip}{0.2cm}
	\begin{align}
	\nonumber E_{fly1}^{\rm UAV}[n] = &P_f
	((x_{j}^{\rm UAV}[n]-x_{j}^{\rm UAV}[n-1])^2+\\&(y_{j}^{\rm UAV}[n]-y_{j}^{\rm UAV}[n-1])^2)^{1/2}V^{-1}.
	\label{thirty-5} 
	\end{align}
}\indent And UAV hovering energy consumption is given by
{
	\setlength{\abovedisplayskip}{0.2cm}
	\setlength{\belowdisplayskip}{0.2cm}
	\begin{align}
	E_{hov1}^{\rm UAV}[n] = P_h(T_{m-j}^{\rm trans}[n] + T_{j-B}^{\rm trans}[n]).
	\label{thirty-6} 
	\end{align} 
}\indent Conclusively, we have $E_{m-j-B}[n]$ as the corresponding total energy consumption of the computing task $U_{m}[n]$, which is denoted as
{
	\setlength{\abovedisplayskip}{0.2cm}
	\begin{align}
	\nonumber &E_{m-j-B}[n] = E_{hov1}^{\rm UAV}[n] + E_{fly1}^{\rm UAV}[n] + E_{m-j}^{\rm trans}[n] \\ &+E_{j-B}^{\rm trans}[n],
	~\forall m\in\mathcal{M},~\forall j\in\mathcal{J^{'}},~\forall n\in\mathcal{N}.
	\label{forty-2} 
	\end{align} 	
}
\vspace{-20pt}
\subsection{Problem Statement}
In this paper, we focus on the problem by jointly optimizing MTUs association, UAV trajectory, transmission power distribution and computation capacity allocation over all time slots where the total energy cost of the entire system is minimized.  
For the sake of simplicity, the MTUs association variable set can be defined as $\textbf{A}$ = $\left\{\alpha_{m}^{j}[n],~\forall m\in\mathcal{M},~\forall n\in\mathcal{N},~\forall j \in \mathcal{J}\right\}$, 
the variable set of UAV trajectory can be defined as $\textbf{U}$ = $\left\{\mathbf{l}_{j}^{\rm UAV}[n],~\forall j \in \mathcal{J'},~\forall n\in\mathcal{N}\right\}$, 
the transmission power distribution variable set can be represented as
$\textbf{P}$ = $\left\{\mathbf{P}_{1},\mathbf{P}_{2},\mathbf{P}_{3}\right\}$,
$\textbf{P}_{1}$ = $\left\{p_{m,j}[n],~\forall m\in\mathcal{M},~\forall j\in\mathcal{K},~\forall n\in\mathcal{N}\right\}$,
$\textbf{P}_{2}$ = $\left\{p_{m,j}^{\rm UAV}[n],~\forall m\in\mathcal{M},~\forall j\in\mathcal{J^{'}},~\forall n\in\mathcal{N}\right\}$,
$\textbf{P}_{3}$ = $\left\{p_{j,B}^{\rm BS}[n],~\forall m\in\mathcal{M},~\forall j\in\mathcal{J^{'}},~\forall n\in\mathcal{N}\right\}$,
and the computation capacity allocation variable set can be expressed as 
$\textbf{F}$ = $\left\{\mathbf{F}_{1},\mathbf{F}_{2},\mathbf{F}_{3}\right\}$,
$\textbf{F}_{1}$ = $\left\{f_{m}^{l}[n],~\forall m\in\mathcal{M},~\forall n\in\mathcal{N}\right\}$,
$\textbf{F}_{2}$ = $\left\{f_{j,m}[n],~\forall m\in\mathcal{M},~\forall j\in\mathcal{K},~\forall n\in\mathcal{N}\right\}$,
$\textbf{F}_{3}$ = $\left\{f_{j,m}^{\rm UAV}[n],~\forall m\in\mathcal{M},~\forall j\in\mathcal{J^{'}},~\forall n\in\mathcal{N}\right\}$.
Under this circumstance, the optimization problem can be represented as
\begin{subequations}
	\begin{align}
	{\bf \mathcal{P}1}:
	&\min_{\textbf{A},\textbf{U},\textbf{P},\textbf{F}}~ \sum_{m=1}^M \sum_{n=1}^{N}E_{m}^{all}[n]  \\ 
	\mbox{s.t.}\quad 
	\label{forty-3a}
	&a_{m}^{j}[n] \in \left\{0,1\right\},~\forall m \in \mathcal{M},~\forall j \in \mathcal{J}, ~\forall n\in\mathcal{N},\\
	\label{forty-3b}
	&\sum_{j=-1}^{Q+K} a_{m}^{j}[n] = 1,~\forall m\in\mathcal{M}, ~\forall n\in\mathcal{N}, \\	
	\label{forty-3c}
	&0 \leq f_{m}^{l}[n] \leq  F_{m,max}^{\rm MTU} , ~\forall m\in\mathcal{M},~\forall n\in\mathcal{N},\\ 
	\label{forty-3d}
	&\nonumber 0 \leq f_{j,m}^{\rm UAV}[n] \leq F_{max}^{\rm UAV},~\forall m\in\mathcal{M},~\forall n\in\mathcal{N},\\&~\forall j\in\mathcal{J^{'}},\\
	\label{forty-3e}
	&\nonumber 0 \leq f_{j,m}[n] \leq F_{j,max}, ~\forall m\in\mathcal{M},~\forall n\in\mathcal{N},\\&~\forall j\in\mathcal{K},\\
	\label{forty-3f} 
	&\nonumber 0 \leq p_{m,j}[n] \leq p_{m,max}, ~\forall m\in\mathcal{M},~\forall n\in\mathcal{N},\\&~\forall j\in\mathcal{K},\\
	\label{forty-3g}
	&\nonumber 0 \leq p_{m,j}^{\rm UAV}[n] \leq p_{m,max}^{\rm UAV},~\forall m\in\mathcal{M},~\forall n\in\mathcal{N},\\&~\forall j\in\mathcal{J^{'}},\\ 
	\label{forty-3h} 
	& 0 \leq p_{j,B}^{\rm BS}[n] \leq p_{max}^{\rm UAV} ,~\forall j\in\mathcal{J^{'}},~\forall n\in\mathcal{N},\\
	\label{forty-3i}
	& a_{m}^{-1}[n]T_{m}^{l}[n] \leq T_m[n] \leq t_m[n], ~\forall m\in\mathcal{M},~\forall n\in\mathcal{N},\\ 
	\label{forty-3j}
	&\nonumber a_{m}^{j}[n](T_{m,j}^{\rm UAV-trans}[n]+T_{m}^{\rm UAV}[n]) \leq T_m[n] \leq t_m[n],\\& ~\forall m\in\mathcal{M},~\forall j\in\mathcal{J^{'}},~\forall n\in\mathcal{N},\\
	\label{forty-3k}
	&\nonumber a_{m}^{j}[n](T_{m,j}[n]+T_{j,m}[n]) \leq T_m[n] \leq t_m[n], ~\forall m\in\mathcal{M},\\&~\forall j\in\mathcal{K},~\forall n\in\mathcal{N},\\
	\label{forty-3l}
	&\nonumber a_{m}^{j}[n](T_{m-j}^{\rm trans}[n] + T_{j-B}^{\rm trans}[n]) \leq T_m[n] \leq t_m[n], \\&~\forall m\in\mathcal{M},~\forall j\in\mathcal{J^{'}},~\forall n\in\mathcal{N},\\
	\label{forty-3m}
	&\nonumber\sum_{n=1}^{N}(a_{m}^{-1}[n]E_{m}^{l}[n]+ \sum_{j=1}^{K}a_{m}^{j}[n]E_{m,j}[n] \\& \nonumber+\sum_{j=K+1}^{K+Q}a_{m}^{j}[n]E_{m,j}^{\rm UAV}[n]  + a_{m}^{0}[n]E_{m-j}^{\rm trans}[n]) \leq  E_{m,max}^{\rm MTU},\\&~\forall m\in\mathcal{M},\\
	\label{forty-3n}
	&\nonumber\sum_{n=1}^{N}\sum_{m=1}^{M}(\sum_{j=K+1}^{K+Q}a_{m}^{j}[n](E_{hov}^{\rm UAV}[n] + E_{fly}^{\rm UAV}[n] + E_{j,m}^{\rm UAV}[n] ))\\& + \nonumber\sum_{n=1}^{N}\sum_{m=1}^{M}a_{m}^{0}[n](E_{hov1}^{\rm UAV}[n] + E_{fly1}^{\rm UAV}[n] + E_{j-B}^{\rm trans}[n]) \\& 
	\sum_{n=1}^{N}\sum_{m=1}^{M}(1-\sum_{j=K+1}^{K+Q}a_{m}^{j}[n]-a_{m}^{0}[n])P_ht_m[n]
	\leq  E_{max}^{\rm UAV},
	\\
	\label{forty-3o}
	&\sum_{n=1}^{N}\sum_{m=1}^{M}a_{m}^{j}[n]E_{j,m}[n]  \leq  E_{j,max},~\forall j\in\mathcal{K},
	\end{align}
	\label{eq1-1}
\end{subequations}

\noindent where $E_{m}^{all}[n]$ illustrates the energy consumption of MTU $m$ to complete the computing task in time slot $n$, 
constraint (\ref{forty-3a}) represents the decision variable of MTUs,
constraint (\ref{forty-3b}) states that each task can only be executed in one place, constraint (\ref{forty-3c})-(\ref{forty-3e}) restrict the maximum CPU frequencies,
constraint (\ref{forty-3f})-(\ref{forty-3h}) ensure that the transmit power cannot exceed the maximum power, constraint (\ref{forty-3i})-(\ref{forty-3l}) indicate that the computing tasks of MTU $m$ in time slot $n$ must be completed within a given period of time $T_m[n]$, and constraint (\ref{forty-3m})-(\ref{forty-3o}) represent the energy constraint of MTU $m$, resource device $j$ and UAV, respectively, which means that the consumed energy to help computing tasks cannot exceed their own energy.

It can be readily observed that the original problem ${\bf \mathcal{P}1}$ is a  mixed integer nonlinear programming problem, which is very complex and hard to be directly solved by conventional optimization techniques. 
In the following section, we resort to decomposing the original problem ${\bf \mathcal{P}1}$ into three more tractable subproblems, namely, MTUs association subproblem, UAV trajectory subproblem, transmission power distribution and computation capacity allocation subproblem.
We design a DRL-based algorithm to achieve a convergent suboptimal solution to the original problem.

\section{Proposed Solution \label{d}}
Due to the nonlinearity of ${\bf \mathcal{P}1}$, it is impractical to come
up with a straight solution. 
Fortunately, the optimal solution to the original problem ${\bf \mathcal{P}1}$ can be found by solving the following three subproblems, i.e., 
we first optimize the $\left\{\textbf{A},\textbf{U}\right\}$ under a given feasible $\left\{\textbf{P},\textbf{F}\right\}$, then we aim to optimize $\textbf{P}$ under the premise of given $\left\{\textbf{A},\textbf{U},\textbf{F}\right\}$, and finally $\textbf{F}$ is optimized when $\left\{\textbf{A},\textbf{U},\textbf{P}\right\}$ are fixed.
In this section, the solutions to these three subproblems are presented, respectively.
\vspace{-10pt}
\subsection{\textbf{MTUs Association and UAV Trajectory Optimization}}

Since the dynamic network environment and system requirements, intelligent approaches are indeed required to achieve better decisions in computation offloading.
As described in \cite{9348616}, Deep Q Network (DQN) approximates the Q-value $Q(s,a)$ by using two deep neural networks (DNNs) with the same several fully connected layers, 
where Q-value refers to the agent at state $s$. After performing the action $a$, the total reward is defined as 
{
	\setlength{\abovedisplayskip}{0.2cm}
	\setlength{\belowdisplayskip}{0.2cm}
	\begin{align}
	Q(s[t],a[t])=\mathbb{E}\big[\sum_{t=0}^{T-1}\omega r[t+1] | s[t], a[t]\big],
	\end{align}
}\noindent where $\omega \in [0,1]$ is the discount factor and $r[t+1]$ is the immediate reward at time $t$ based on the state-action pair $(s[t],a[t])$. The meaning of $Q(s[t],a[t])$ is to evaluate how well the agent performs action $a[t]$ in state $s[t]$.
Unfortunately, the DQN algorithm may lead to the over-estimation of Q-value. To solve this problem, the Double DQN (DDQN) algorithm tries to avoid over-estimation by separating the selection action from the evaluation action. 
In this subsection, we use DDQN algorithm to explore the unknown environment and optimize MTUs association and UAV trajectory.
\\
\indent From Eqs. (\ref{four}) and (\ref{five}), the location of MTUs possess Markov characteristics, which makes the whole process follow the MDP.
As such, for any given transmission power distribution $\textbf{P}$ and computation capacity allocation $\textbf{F}$, the MTUs association and UAV trajectory of problem ${\bf \mathcal{P}1}$ can be optimized by DRL. 
In this subsection, we first elaborate four key elements for RL, and then we use DDQN algorithm to explore the unknown environment and optimize MTUs association and UAV trajectory, which not only solves the problem of overestimation of DQN, but also solves the problem of large number of state-action pairs caused by MTUs position change. 
\begin{itemize}
	\item
	\textbf{Four Key Elements for RL:}
	There are four key elements in the RL method, namely agent \& environment, state, action and reward, specifically to the system model in this paper.
	
	$\mathbf{Agent ~\&~Environment}$: In our proposed DITEN system, the goal of the agent in the environment is to maximize its potential future reward. 
	Hence, different from other RL methods, our model transforms the minimal sum energy consumption to maximal reward by defining the reward negatively correlated with energy cost.
	
	$\mathbf{State}$: The system state consists of three components
	{
		\setlength{\abovedisplayskip}{0.2cm}
		\setlength{\belowdisplayskip}{0.2cm}
		\begin{align}
		 S = \left \{s[n] | s[n] = \{\mathbf{l}_{m}^{\rm MTU}[n], \mathbf{l}_{j}^{\rm UAV}[n], U_m[n]\right \},
		\end{align} 
	}\noindent where $\mathbf{l}_{m}^{\rm MTU}[n]$ and $\mathbf{l}_{j}^{\rm UAV}[n]$ response to the MTU $m$ and UAV location in time slot $n$, respectively, and $ U_m[n]$ is the information about the task generated by MTU $m$ in time slot $n$.
	The agent transitions will from state to other specific state after executing an action.
	\begin{figure}[t]
		\flushleft
		\includegraphics[width=3.6in]{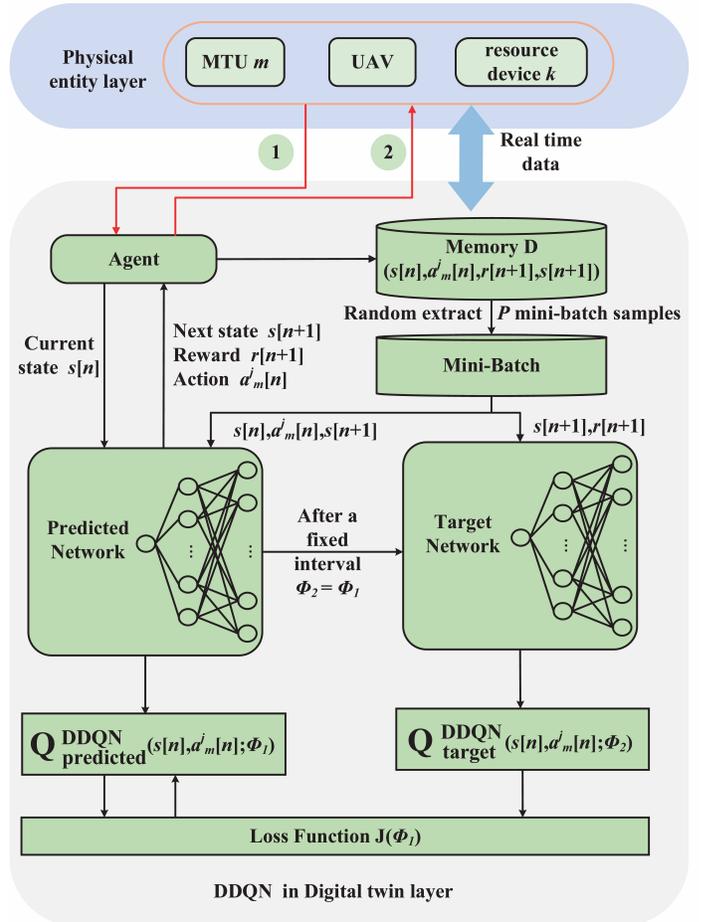}
		\caption{The DDQN training with DT assistance.}
		\label{DT-DDQN}
	\end{figure}
	
		$\mathbf{Action}$: Synthesizing our proposed MEC network model, the action consists of the following part
		{
			\setlength{\abovedisplayskip}{0.2cm}
			\setlength{\belowdisplayskip}{0.2cm}
			\begin{align}
			A = \left\{a_{m}^{j}[n],~\forall m \in \mathcal{M},~\forall j \in \mathcal{J},~\forall n \in \mathcal{N}\right\},
			\end{align}
		}\noindent where $a_{m}^{j}[n]$ represents MTU $m$'s offloading decision in the $n$-th time slot. By performing an action, the agent transits from state to next state.

		$\mathbf{Reward}$: At every step, the agent will get a reward $r[n+1]$
		in the certain state $s[n]$ after executing each possible action $a_{m}^{j}[n]$.
		In a sense, the reward function should be related to the objective function. However, our objective function is to minimize the total energy consumption of the system, the goal of the RL is to maximize the reward. To this end, the value of the reward should be negatively correlated with the objective function and we define the immediate reward as  
		{
			\setlength{\abovedisplayskip}{0.2cm}
			\setlength{\belowdisplayskip}{0.2cm}
			\begin{align}
			\nonumber &r[n+1] = - \sum_{m=1}^M E_{m}^{all}[n]-\vartheta,
			\end{align}
		}\noindent where $\vartheta$ is the penalty if any MTU executes the task beyond the maximum processing delay, which means that one of contraints (\ref{forty-3i}), (\ref{forty-3j}), (\ref{forty-3k}) and (\ref{forty-3l}) is not satisfied.
\end{itemize}
\begin{itemize}
	\item
		\setlength{\belowdisplayskip}{-2cm}
	\textbf{MTUs Association and UAV Trajectory Optimization:}
	Given the real-time locations of a set of MTUs, the transmission power distribution $\mathbf{P}$, and the computation capacity allocation $\mathbf{F}$, the corresponding real-time MTUs association $\textbf{A}$ and UAV trajectory $\textbf{U}$ optimization problem can be formulated as
	\begin{subequations}
		\setlength{\abovedisplayskip}{0.2cm}
		\setlength{\belowdisplayskip}{0.2cm}
		\begin{align}
		{\bf \mathcal{P}1.1}:
		&\argmax_{\pi^{*}}~ \sum_{n=0}^{N-1}r[n+1]\\ 
		\mbox{s.t.}\quad 
		&(\rm \ref{forty-3a}),(\ref{forty-3b}),(\ref{forty-3i})-(\ref{forty-3o}),
		\end{align} 
	\end{subequations}\noindent where $\pi^{*}$ stands for the optimal strategy of $\left\{\textbf{A},\textbf{U}\right\}$.
		
	\setlength\parindent{1em}To solve the MDP problem ${\bf \mathcal{P}1.1}$, we use DDQN with experience replay optimization algorithm to obtain the optimal strategy $\pi^{*}$.
	As shown in Fig. \ref{DT-DDQN}, DT maps all aspects of the physical objects MTU, UAV and resource device in the entire system environment to virtual space in real time, forming a digital mirror image. 
	At the same time, the DRL agent interacts with the DT of the physical objects to learn and obtains the optimal strategy $\pi^{*}$.
	For the target network, the input is still the next state $s[n+1]$, but the action is not selected according to the maximum Q-value. In fact, the corresponding action is obtained through the predicted network, i.e., $\mathop{\argmax}\limits_{a'}Q(s[n+1],a';\phi_1)$, and then the corresponding Q-value of next state-action pair is obtained, i.e., $Q(s[n+1], \mathop{\argmax}\limits_{a'}Q(s[n+1],a';\phi_1);\phi_2)$.
	Therefore, the target network value in DDQN is denoted as 
	{
		\setlength{\abovedisplayskip}{0.2cm}
		\setlength{\belowdisplayskip}{0.2cm}
		\begin{align}
		\nonumber Q_{target}^{DDQN}(&s[n],a_{m}^{j}[n];\phi_2)=r[n+1] + \omega Q \big(s[n+1], \\ &\argmax_{a'}Q(s[n+1],a';\phi_1);\phi_2\big).
		\label{fifty-8}
		\end{align}
	}\indent Through DT, the agent achieves the same training effect with the real environment at a lower cost.
	
	\setlength\parindent{1em}The details of DDQN with experience replay algorithm are summarized in Algorithm $1$.
	From Algorithm $1$, we can see that MTU $m$ uses conventional $\epsilon$-greedy policy to select a random action $a_{m}^{j}[n]$ from action space $A$ with probability $\epsilon$ and $a_{m}^{j}[n] =
	\mathop{\argmax}\limits_{a'}Q(s[n],a';\phi_{1})$ with probability $(1-\epsilon)$ based on the current state $s[n]$.
	Based on the action taken by the agent, the state of the system changes to a new state $s[n+1]$ and receives a reward $r[n+1]$ which is determined by the instantaneous energy consumption of the entire system. Meanwhile, the transition tuple $(s[n],a_{m}^{j}[n],r[n+1],s[n+1])$ is collected and stored into the replay memory D with a size of $Z$.
	Then, if the memory D is full, $P$ mini-batch samples are randomly extracted from D to update $\phi_{1}$, the loss function can be represented as
	{	
		\setlength{\abovedisplayskip}{0.2cm}
		\setlength{\belowdisplayskip}{0.2cm}
		\begin{align}
		J(\phi_{1})=\frac{1}{P}\sum_{p=1}^{P}[Q_{target}^{DDQN}(p)-Q_{predicted}^{DDQN}(p)]^{2},
		\end{align}
	}\noindent where $Q_{target}^{DDQN}(p)$ and $Q_{predicted}^{DDQN}(p)$ represent the target and predicted values of the $p$-th sample from the $P$ mini-batch samples, repectively. It is clear that the gradient descent method is applied to update $\phi_{1}$ of the predicted network as 
    {	
    	\setlength{\abovedisplayskip}{0.2cm}
    	\setlength{\belowdisplayskip}{0.2cm}
    	\begin{align}
    	\phi_{1} = \phi_{1} - \lambda \nabla_{\phi_{1}}J(\phi_{1}),
    	\end{align}
    }\noindent where $\lambda$ is the learning rate that satisfies $0 \leq \lambda \leq 1$
	and $\nabla_{\phi_{1}}$ is the gradient function with respect to $\phi_{1}$.
	Moreover, $\phi_{2}$ is updated as $\phi_{2} = \phi_{1}$ after a fixed interval.
	To achieve a good tradeoff between exploration and exploitation, a decrement $\theta$ is subtracted from $\epsilon$ in line $18$. Finally, the proposed algorithm produces the optimal policy $\pi^{*}$.
\end{itemize}

\subsection{\textbf{Transmission Power Distribution Optimization}}
In this subsection, we optimize the transmission power distribution $\textbf{P}$. 
We can get the following optimization problem
{
	\setlength{\belowdisplayskip}{0.2cm}
	\setlength{\abovedisplayskip}{0.2cm}
	\begin{subequations}
		\begin{align}
		{\bf \mathcal{P}1.2}:
		&\min_{\textbf{P}}~ \sum_{m=1}^M \sum_{n=1}^{N}E_{m}^{all}[n]  \\ 
		\mbox{s.t.}\quad 	
		&(\rm \ref{forty-3f})-(\ref{forty-3h}), (\ref{forty-3j})-(\ref{forty-3o}).
		\end{align}
		\label{eq1-4}
	\end{subequations}
}\indent \textit{Proposition 1:} The energy consumption of MTU $m$ associated
with resource device $j$ is increasing with respect to the transmit power $p_{m,j}[n]$.

\textit{Proof:} The energy consumption of MTU $m$ associated
with resource device $j$ is given by
{	
	\setlength{\abovedisplayskip}{0.2cm}
	\setlength{\belowdisplayskip}{0.2cm}
	\begin{align}
	\nonumber E_{m,j}[n] &= p_{m,j}[n] T_{m,j}[n] 
	=p_{m,j}[n]\frac{D_{m}[n]}{R_{m,j}[n]} \\
	& \nonumber =p_{m,j}[n]\frac{D_{m}[n]}{ B\log_2(1 + \frac{p_{m,j}[n] h_{m,j}[n]}{\sigma^2})},   
	~\forall m\in\mathcal{M},\\&~\forall j\in\mathcal{K},~\forall n\in\mathcal{N}.
	\label{E-rd} 
	\end{align} 
}\indent The first-order derivative of $E_{m,j}[n]$ with respect to $p_{m,j}[n]$ is denoted as
{	
	\setlength{\abovedisplayskip}{0.2cm}
	\setlength{\belowdisplayskip}{0.2cm}
	\begin{align}
	\nonumber &\frac{\partial E_{m,j}[n]}{\partial p_{m,j}[n]} 
	=(D_{m}[n]\ln 2[\ln(1+\frac{p_{m,j}[n] h_{m,j}[n]}{\sigma^2})-\\& \nonumber \frac{p_{m,j}[n] h_{m,j}[n]}{p_{m,j}[n] h_{m,j}^{\rm RD}[n]+\sigma^2}])
	(B[\ln(1 + \frac{p_{m,j}[n] h_{m,j}[n]}{\sigma^2})]^{2})^{-1},   
	\\&~\forall m\in\mathcal{M},~\forall j\in\mathcal{K},~\forall n\in\mathcal{N}.
	\label{E-rd1}  	
	\end{align}
}\indent Afterwards, we define $x$ as following
{	
	\setlength{\abovedisplayskip}{0.2cm}
	\setlength{\belowdisplayskip}{0.2cm}
	\begin{align}
	\nonumber x = 1+p_{m,j}[n] &h_{m,j}[n](\sigma^2)^{-1} =(p_{m,j}[n] h_{m,j}[n]+\sigma^2)\\&(\sigma^2)^{-1},
	~\forall m\in\mathcal{M},~\forall j\in\mathcal{K},~\forall n\in\mathcal{N}.
	\label{x}  
	\end{align}
}\indent From $(\ref{x})$, we have $x \geq 1$ and $1-{x}^{-1} = (p_{m,j}[n] h_{m,j}[n])(p_{m,j}[n] h_{m,j}[n]+\sigma^2)^{-1}$. Consequently,
the numerator of $(\ref{E-rd1})$ can be expressed as 
{   \setlength{\abovedisplayskip}{0.2cm}
	\setlength{\belowdisplayskip}{0.2cm}
	\begin{align}
	g(x) = \ln(x)-(1-\frac{1}{x}),
	\label{E-rd2}
	\end{align}
}and its first-order derivative is given by 
{
	\setlength{\abovedisplayskip}{0.2cm}
	\setlength{\belowdisplayskip}{0.2cm}
	\begin{align}
	\frac{\partial g(x)}{\partial x} = \frac{1}{x}-\frac{1}{x^2} = \frac{x-1}{x^2} \geq 0.
	\label{E-rd3} 
	\end{align}
}\indent Thus, we have $g(x) \geq 0$ and ${\partial E_{m,j}[n]}{(\partial p_{m,j}[n]})^{-1} \geq 0$, which indicates that $E_{m,j}[n]$ is
nondecreasing with $p_{m,j}[n]$.

Since the energy consumption of MTU $m$ increases with the MTU transmit power $p_{m,j}[n]$, the optimal transmit power of MTU $m$ to resource device $j$ can be obtained by satisfying the latency constraints.

\textit{Theorem 1:} The optimal transmit power of MTU $m$ to resource device $j$ can be concluded as
{
	\setlength{\abovedisplayskip}{0.2cm}
	\setlength{\belowdisplayskip}{0.2cm}
	\begin{align}
	\nonumber (p_{m,j}[n])^{*} =& \rm min\left\{\xi_{m,j}[n]\frac{\sigma^2}{h_{m,j}[n]},
	p_{m,max}\right\},\\&~\forall m\in\mathcal{M},~\forall n\in\mathcal{N},~\forall j\in\mathcal{K},
	\end{align}	
}\noindent where  $\xi_{m,j}[n] = 2^{\frac{D_{m}[n]}{B(T_m[n] - \Delta  T_{j,m}[n] - \widetilde{T}_{j,m}[n])}} - 1$.

\textit{Proof 1:} To minimize the power consumption of MTU $m$ associated with resource device $j$, the latency constraint should be satisfied, i.e., $T_{m,j}[n] +  T_{m}[n] = \frac{D_{m}[n]}{R_{m,j}[n]}+
\Delta  T_{j,m}[n] + \widetilde{T}_{j,m}[n]=T_m[n]$, and we have
{
	\setlength{\abovedisplayskip}{0.2cm}
	\setlength{\belowdisplayskip}{0.2cm}
	\begin{align}
	B\log_2(1 + \frac{p_{m,j}[n] h_{m,j}[n]}{\sigma^2})
	= \frac{D_{m}[n]}{T_m[n] - \Delta  T_{j,m}[n] - \widetilde{T}_{j,m}[n]}.
	\label{E-rd4}
	\end{align}
}\indent By calculating the transmit power $p_{m,j}[n]$ in (\ref{E-rd4}), we get
{
	\setlength{\abovedisplayskip}{0.2cm}
	\setlength{\belowdisplayskip}{0.2cm}
	\begin{align}
	p_{m,j}[n] = (2^{\frac{D_{m}[n]}{B(T_m[n]-\Delta  T_{j,m}[n] - \widetilde{T}_{j,m}[n])}}-1) \frac{\sigma^2}{h_{m,j}[n]}.
	\end{align}
}\indent As a result, we obtain the closed-form expressions as shown in $\textit{Theorem 1}$, which completes the proof.

\textit{Proposition 2:} The energy consumption of MTU $m$ associated
with UAV $j$ is increasing with respect to the transmit power $p_{m,j}^{\rm UAV}[n]$.

The detailed proof of $\textit{Proposition 2}$ can be seen in Appendix A.

\textit{Theorem 2:} The optimal transmit power of MTU $m$ to UAV $j$ can be described as
{
	\setlength{\abovedisplayskip}{0.2cm}
	\setlength{\belowdisplayskip}{0.2cm}
	\begin{align}
	\nonumber (p_{m,j}^{\rm UAV}[n])^{*} =& \rm min\left\{\xi_{m,j}^{\rm UAV}[n]\frac{\sigma^2}{h_{m,j}^{\rm UAV}[n]},
	p_{m,max}^{\rm UAV}\right\},\\&~\forall m\in\mathcal{M},~\forall n\in\mathcal{N},~\forall j\in\mathcal{J^{'}},
	\end{align}
}\noindent where  $\xi_{m,j}^{\rm UAV}[n] = 2^{\frac{D_{m}[n]}{B(T_m[n] - \Delta  T_{j,m}^{\rm UAV}[n] - \widetilde{T}_{j,m}^{\rm UAV}[n])}}-1$.

\textit{Proof 2:} To minimize the power consumption of MTU $m$ associated with UAV $j$, the latency constraint should be satisfied, i.e., $T_{m,j}^{\rm {UAV}-trans}[n] + T_{m}^{\rm UAV}[n] = \frac{D_{m}[n]}{R_{m,j}^{\rm UAV}[n]}+
\widetilde{T}_{j,m}^{\rm UAV}[n] + \Delta T_{j,m}^{\rm UAV}[n] = T_m[n]$, and we have
{
	\setlength{\abovedisplayskip}{0.2cm}
	\setlength{\belowdisplayskip}{0.2cm}
	\begin{small}
		\begin{align}
		B\log_2(1 + \frac{p_{m,j}^{\rm UAV}[n] h_{m,j}^{\rm UAV}[n]}{\sigma^2})
		= \frac{D_{m}[n]}{T_m[n] - \widetilde{T}_{j,m}^{\rm UAV}[n] - \Delta T_{j,m}^{\rm UAV}[n]}.
		\label{E-rd9}
		\end{align}
	\end{small}
}\indent By calculating the transmit power $p_{m,j}^{\rm UAV}[n]$ in (\ref{E-rd9}), we get
{
	\setlength{\abovedisplayskip}{0.2cm}
	\setlength{\belowdisplayskip}{0.2cm}
	\begin{align}
	p_{m,j}^{\rm UAV}[n] = (2^{\frac{D_{m}[n]}{B(T_m[n]-\widetilde{T}_{j,m}^{\rm UAV}[n] - \Delta T_{j,m}^{\rm UAV}[n])}}-1) \frac{\sigma^2}{h_{m,j}^{\rm UAV}[n]}.
	\end{align}
}\indent Finally, we obtain the closed-form expressions as shown in $\textit{Theorem 2}$, which completes the proof.
\begin{figure}[t]
	\includegraphics[width=3.6in]{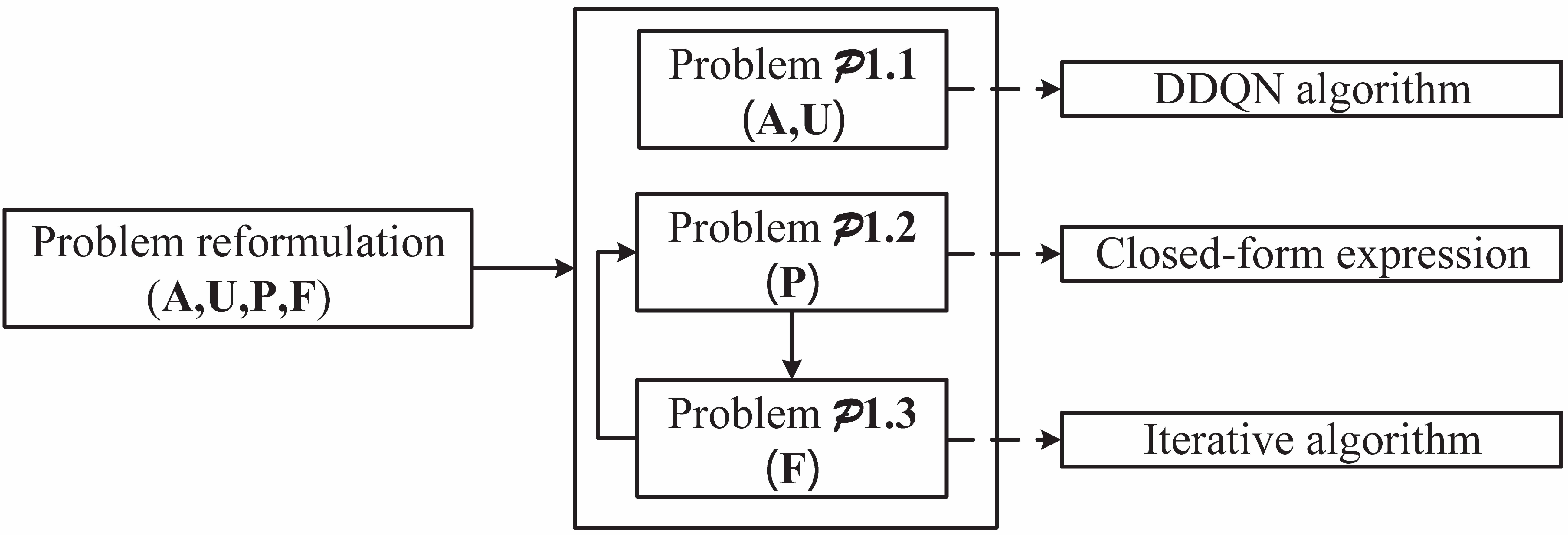}
	\caption{The procedure of Algorithm 2.}
	\label{Algorithm}
\end{figure}

\textit{Proposition 3:} The energy consumption of UAV $j$ associated
with BS is increasing with respect to the transmit power $p_{j,B}^{\rm BS}[n]$.

The detailed proof of $\textit{Proposition 3}$ is similar to Appendix A and is omitted due to the space limitation.

\textit{Theorem 3:} The optimal transmit power of UAV $j$ to BS can be represented as
{
	\setlength{\abovedisplayskip}{0.2cm}
	\setlength{\belowdisplayskip}{0.2cm}
	\begin{align}
	\nonumber (p_{j,B}^{\rm BS}[n])^{*} =& \rm min\left\{\xi_{j,B}^{\rm BS}[n]\frac{\sigma^2}{h_{j-B}[n]},
	p_{max}^{\rm UAV}\right\},\\&~\forall m\in\mathcal{M},~\forall n\in\mathcal{N},~\forall j\in\mathcal{J^{'}},
	\end{align}	
}\noindent where  $\xi_{m,j}^{\rm BS}[n] = 2^{\frac{D_{m}[n]}{B(T_m[n]-T_{m-j}^{\rm trans}[n])}}-1$.

\textit{Proof 3:} To minimize the power consumption of UAV $j$ associated with BS, the latency constraint should be satisfied, i.e., $T_{m-j}^{\rm trans}[n] +
T_{j-B}^{\rm trans}[n] = T_{m-j}^{\rm trans}[n] + \frac{D_{m}[n]}{R_{j-B}^{\rm trans}[n]} = T_m[n]$, and we have
{
	\setlength{\abovedisplayskip}{0.2cm}
	\setlength{\belowdisplayskip}{0.2cm}
	\begin{align}
	B\log_2(1 + \frac{p_{j,B}^{\rm BS}[n] h_{j-B}[n]}{\sigma^2})
	= \frac{D_{m}[n]}{T_m[n] - T_{m-j}^{\rm trans}[n]}.
	\label{E-rd10}
	\end{align}
}\indent By calculating the transmit power $p_{j,B}^{\rm BS}[n]$ in (\ref{E-rd10}), we get
{
	\setlength{\abovedisplayskip}{0.2cm}
	\setlength{\belowdisplayskip}{0.2cm}
	\begin{align}
	p_{j,B}^{\rm BS}[n] = (2^{\frac{D_{m}[n]}{B(T_m[n]-T_{m-j}^{\rm trans}[n])}}-1) \frac{\sigma^2}{h_{j-B}[n]}.
	\end{align}
}\indent Consequently, we obtain the closed-form expressions as shown in $\textit{Theorem 3}$, which completes the proof.
\vspace{-10pt}
\subsection{\textbf{Computation Capacity Allocation Optimization}}
In this subsection, we optimize the computation capacity allocation $\textbf{F}$. 
We can get the following optimization problem
\begin{algorithm}[t]
	\caption{The DDQN with Experience Replay}
	
	\noindent \hangafter=1 \setlength{\hangindent}{1.8em}
	\hspace{-0.1em}\textbf{1:~}$\mathbf{Input}$:
	$\phi_{1},\phi_{2},\epsilon,\theta,P,\lambda,Number_{e},N$;
	
	\noindent \hangafter=1 \setlength{\hangindent}{1.8em}
	\hspace{-0.1em}\textbf{2:~}$\mathbf{for}$
	episode = $1$ to $Number_{e}$ $\mathbf{do}$
	
	\noindent \hangafter=1 \setlength{\hangindent}{1.8em}
	\hspace{-0.1em}\textbf{3:~~~}$\mathbf{for}$
	 n = $1$ to $N$ $\mathbf{do}$
	 
	 \noindent \hangafter=1 \setlength{\hangindent}{1.8em}
	 \hspace{-0.1em}\textbf{4:~~~~}
	 Get the initial state $s[n]$;
	 
	\noindent \hangafter=1 \setlength{\hangindent}{1.8em}
	\hspace{-0.1em}\textbf{5:~~~~}
	Take action $a_{m}^{j}[n]$ with $\epsilon$-greedy policy at $s[n]$;
	
	\noindent \hangafter=1 \setlength{\hangindent}{1.8em}
	\hspace{-0.1em}\textbf{6:~~~~}
	Case $\rm\uppercase\expandafter{\romannumeral1}$: with probability $\epsilon$ select a random action $a_{m}^{j}[n]$;
	
	\noindent \hangafter=1 \setlength{\hangindent}{1.8em}
	\hspace{-0.1em}\textbf{7:~~~~}
	Case $\rm\uppercase\expandafter{\romannumeral2}$: select $a_{m}^{j}[n]=\mathop{\argmax}\limits_{a'}Q(s[n],a';\phi_1)$;

	\noindent \hangafter=1 \setlength{\hangindent}{1.8em}
	\hspace{-0.1em}\textbf{8:~~~~}
	Obtain the reward $r[n+1]$ and transfer to $s[n+1]$;
	
	\noindent \hangafter=1 \setlength{\hangindent}{1.8em}
	\hspace{-0.1em}\textbf{9:~~~~}
	Store the transition $(s[n],a_{m}^{j}[n],r[n+1],s[n+1])$ 
	in the memory D with a size of $Z$;
	
	\noindent \hangafter=1 \setlength{\hangindent}{1.8em}
	\hspace{-0.1em}\textbf{10:~~~~}
	$\mathbf{if}$ D is full $\mathbf{then}$
	
	\noindent \hangafter=1 \setlength{\hangindent}{1.8em}
	\hspace{-0.1em}\textbf{11:~~~~~}
	Random extract $P$ mini-batch samples from D;
	
	\noindent \hangafter=1 \setlength{\hangindent}{1.8em}
	\hspace{-0.1em}\textbf{12:~~~~~}
	$\mathbf{for}$ $p=1$ to $P$ $\mathbf{do}$
	
	\noindent \hangafter=1 \setlength{\hangindent}{1.8em}
	\hspace{-0.1em}\textbf{13:~~~~~~}
	Obtain $Q_{predicted}^{DDQN}(p)$ and $Q_{target}^{DDQN}(p)$;
	
	\noindent \hangafter=1 \setlength{\hangindent}{1.8em}
	\hspace{-0.1em}\textbf{14:~~~~~}
	$\mathbf{end}$ $\mathbf{for}$
	
	\noindent \hangafter=1 \setlength{\hangindent}{1.8em}
	\hspace{-0.1em}\textbf{15:~~~~~~}
	Update $\phi_{1}$ with $\phi_{1} = \phi_{1} - \lambda \nabla_{\phi_{1}}J(\phi_{1})$;
	
	\noindent \hangafter=1 \setlength{\hangindent}{1.8em}
	\hspace{-0.1em}\textbf{16:~~~~~~}	
	After a fixed interval, update $\phi_{2}$ as $\phi_{2} = \phi_{1}$;
	
	\noindent \hangafter=1 \setlength{\hangindent}{1.8em}
	\hspace{-0.1em}\textbf{17:~~~~~}		
	$\mathbf{end}$ $\mathbf{if}$

	\noindent \hangafter=1 \setlength{\hangindent}{1.8em}
	\hspace{-0.1em}\textbf{18:~~}
	$n=n+1$ and $\epsilon = \epsilon - \theta$;
	
	\noindent \hangafter=1 \setlength{\hangindent}{1.8em}
	\hspace{-0.1em}\textbf{19:~~}
	$\mathbf{end}$ $\mathbf{for}$

	\noindent \hangafter=1 \setlength{\hangindent}{1.8em}
	\hspace{-0.1em}\textbf{20:}
    $\mathbf{end}$ $\mathbf{for}$
    
    \noindent \hangafter=1 \setlength{\hangindent}{1.8em}
    \hspace{-0.1em}\textbf{21:}
    $\mathbf{Output}$: The optimal policy $\pi^{*}$.
\end{algorithm}

\begin{subequations}
	\begin{align}
	{\bf \mathcal{P}1.3}:
	&\min_{\textbf{F}}~ \sum_{m=1}^M \sum_{n=1}^{N}E_{m}^{all}[n]  \\ 
	\mbox{s.t.}\quad 	
	&(\rm \ref{forty-3c})-(\ref{forty-3e}), (\ref{forty-3i})-(\ref{forty-3o}).
	\end{align}
	\label{eq1-3}
\end{subequations}

It is noteworthy that ${\bf \mathcal{P}1.3}$ is a standard linear program
problem when the optimal policy $\pi^{*}$ and transmission power distribution $\textbf{P}$ are known, so we can use the optimization tool (such as CVX) to
solve it effectively in an iterative way.
\vspace{-10pt}
\subsection{\textbf{Joint Algorithm Design}}
Based on the results presented in the previous three subsections, we propose a joint algorithm for original problem ${\bf \mathcal{P}1}$.
The detailed steps are summarized in Fig. \ref{Algorithm}. Specifically, 
the entire optimization variables in original problem ${\bf \mathcal{P}1}$ are partitioned into three parts, i.e., $\left\{\mathbf{A},\mathbf{U}\right\}$, $\mathbf{P}$ and $\mathbf{F}$.
Then, the MTUs’ association $\mathbf{A}$, the UAV trajectory $\textbf{U}$, the
transmission power distribution $\textbf{P}$ and computation capacity allocation $\mathbf{F}$ are optimized by solving problem ${\bf \mathcal{P}1.1}$, Theorems $1$, $2$, $3$ and ${\bf \mathcal{P}1.3}$, respectively.
The details of this algorithm are summarized in Algorithm $2$.
\begin{algorithm}[t]
	\caption{Joint optimization algorithm for MTUs association, UAV trajectory, transmission power distribution and computation capacity allocation}
	
	\hangafter=1 \setlength{\hangindent}{1.8em}
	\hspace{-0.1em}\textbf{1:}$\mathbf{Input}$:
	 the locations of $Q$ FHPs, $K$ resource devices and a BS,
	 $\phi_{1},\phi_{2},\epsilon,\theta,P,\lambda,Number_{e},N,\varsigma$, let $r$ = 0;
	
	\hangafter=1 \setlength{\hangindent}{1.8em}
	\hspace{-0.1em}\textbf{2:~}Solve problem ${\bf \mathcal{P}1.1}$ for given ${\mathbf{P}}$ and $\mathbf{F}$, and denote the optimal solution as $\pi^{*}$;
	
	\hangafter=1 \setlength{\hangindent}{1.8em}
	\hspace{-0.1em}\textbf{3:~}$\mathbf{repeat}$ 
	
	\hangafter=1 \setlength{\hangindent}{1.8em}
	\hspace{-0.1em}\textbf{4:~~}~~Obtain $p_{m,j}[n]$, $p_{m,j}^{\rm UAV}[n]$ and $p_{j,B}^{\rm BS}[n]$ in closed-form based on Theorem $1$, $2$ and $3$, respectively, and denote the optimal solution as $\left\{\mathbf{P}^{r}\right\}$;
		
	\hangafter=1 \setlength{\hangindent}{1.8em}
	\hspace{-0.1em}\textbf{5:~~}~~Solve problem ${\bf \mathcal{P}1.3}$ for given $\pi^{*}$ and $\left\{\mathbf{P}^{r}\right\}$, and denote the optimal solution as $\left\{\mathbf{F}^{r}\right\}$;
	
	\hangafter=1 \setlength{\hangindent}{1.8em}
	\hspace{-0.1em}\textbf{6:~~~}~~Update r = r + 1;
	
	\hangafter=1 \setlength{\hangindent}{1.8em}
	\hspace{-0.1em}\textbf{7:~}$\mathbf{until}$~~the fractional increase of the objective value is below a threshold $\varsigma \textgreater0$;
	
	\hangafter=1 \setlength{\hangindent}{1.8em}
	\hspace{-0.1em}\textbf{8:}
	$\mathbf{Output}$: The MTUs association $\mathbf{A}$, UAV trajectory $\mathbf{U}$, transmission power distribution $\mathbf{P}$, and computation capacity allocation $\mathbf{F}$.
\end{algorithm}
\section{Complexity and Convergence Analysis \label{e}}
In this section, we first give a brief analysis of the computational complexity of Algorithm $2$.

\textit{Complexity Analysis:}
In order to solve the problem ${\bf \mathcal{P}1.1}$, we adopt the DDQN method, which involves the neural network.
It is well known that the computational complexity of neural networks is affected by many factors, such as the size of data, the complexity of models and the overall algorithm framework.  
It is very complicated to analyze the complexity of neural networks, and few research dealt with such a problem.  
In order to simplify this problem, we focus on the computational complexity of generating optimal actions.  
In each iteration, each agent in the DDQN traverses all actions to find the optimal action with the maximal Q value. In our proposed MEC system, there are $M$ MTUs in each time slot, and each MTU can choose one from $(K+Q+2)$ actions. Therefore, the corresponding computational complexity is 
{
	\setlength{\abovedisplayskip}{0.2cm}
	\setlength{\belowdisplayskip}{0.2cm}
	\begin{align}
	\mathcal{O}[NM(K+Q+2)],
	\label{sixty-3}
	\end{align} 
}and the complexity to solve the computation capacity allocation of problem ${\bf \mathcal{P}1.3}$ is $\mathcal{O}(EN(M+Q+K))$, where $E$ represents the number of external iterations.
Then we can get the total complexity of Algorithm $2$ as follows
{
	\setlength{\abovedisplayskip}{0.2cm}
	\setlength{\belowdisplayskip}{0.2cm}
	\begin{align}
	\mathcal{O}[NM(K+Q+2) +EN(M+Q+K)].
	\label{sixty-4}
	\end{align}
}\indent \textit{Convergence Analysis:} First, we analyze the convergence of DDQN algorithm under different learning rates in Fig. \ref{Compare}. It can be observed that the higher the learning rate is, the faster the convergence speed of DDQN is.
In addition, it can be seen that the learning rate increases, we suffer from a larger possibility of obtaining a local optimal solution instead of the global optimal one. Hence, we need to choose an appropriate learning rate regarding to specific situations.

On the other hand, it is noted that the transmission power distribution problem and the computation capacity allocation problem can only optimally solve its approximation problem. 
Therefore, we need to prove the convergence, rather than directly apply
the correlation convergence analysis of classical algorithm.
For the objective value of the original problem ${\bf \mathcal{P}1}$, we define as
$\Phi(\textbf{A},\textbf{U},\textbf{P},\textbf{F})$. 
In step $4$ of Algorithm $2$, since $\left\{\mathbf{P}^{r+1}\right\}$
is one subopimal transmission power distribution with the fixed $\left\{\mathbf{F}^{r}\right\}$, we have
{	
	\setlength{\abovedisplayskip}{0.2cm}
	\setlength{\belowdisplayskip}{0.2cm}
	\begin{align}
	\Phi(\textbf{A},\textbf{U},\textbf{P}^{r},\textbf{F}^{r}) \geq \Phi(\textbf{A},\textbf{U},\textbf{P}^{r+1},\textbf{F}^{r}).
	\label{shoulian-1}
	\end{align} 
	
}\indent In step $5$ of Algorithm $2$, for given $\Phi(\textbf{A},\textbf{U},\textbf{P}^{r+1},\textbf{F}^{r})$, $\left\{\mathbf{F}^{r+1}\right\}$ is one subopimal computation capacity allocation of problem ${\bf \mathcal{P}1.3}$, it follows that
{	
	\setlength{\abovedisplayskip}{0.2cm}
	\setlength{\belowdisplayskip}{0.2cm}
	\begin{align}
	\Phi(\textbf{A},\textbf{U},\textbf{P}^{r+1},\textbf{F}^{r}) \geq \Phi(\textbf{A},\textbf{U},\textbf{P}^{r+1},\textbf{F}^{r+1}).
	\label{shoulian-2}
	\end{align} 
}\indent Based on the above analysis, we can get
{	
	\setlength{\abovedisplayskip}{0.2cm}
	\setlength{\belowdisplayskip}{0.2cm}
	\begin{align}
	\Phi(\textbf{A},\textbf{U},\textbf{P}^{r},\textbf{F}^{r}) \geq \Phi(\textbf{A},\textbf{U},\textbf{P}^{r+1},\textbf{F}^{r+1}),
	\label{shpulian-3}
	\end{align} 
}which indicates that after each iteration of Algorithm $2$, the target value of problem ${\bf \mathcal{P}1}$ is nonincreasing. Since the target value of problem ${\bf \mathcal{P}1}$ is the lower bound of a finite value, the proposed Algorithm $2$ can guarantee convergence. 
\section{Numerical Results \label{f}}
In this section, we will show the performance of our proposed architecture for DITEN with numerical results.
We consider a system with $K$ = 10 and $Q$ = 15. The UAV is assumed to fly at a fixed altitude $H = 500~\rm m$.
The maximum CPU computation capacity of MTUs, resource devices and UAV are assumed as $F_{m,max}^{\rm MTU} = 6~\rm {GHz}$, $F_{j,max} = 8~\rm {GHz}$ and  $F_{max}^{\rm UAV} = 10~\rm {GHz}$, respectively.
The powers of UAV flying and hovering are set to $P_f=0.11~\rm W$ and $P_h=0.08~\rm W$, respectively. Some other parameters {\cite{9244624}, \cite{9399641}, \cite{9311405}, \cite{9162853}} are listed in Table $\rm{I}$.

\begin{table}[t]	
	\caption{PARAMETER VALUE SETTING}	
	\begin{center}
		\begin{tabu} to 0.5\textwidth{X[2,c]|X[2,c]|X[2,c]|X[3,c]}
			\hline
			$\rm Parameter$      &$\rm Value$      &$\rm Parameter$      &$\rm Value$    \\ 
			\hline
			\rule{0pt}{10pt} 
			$\epsilon$    &0.95    &$N$           & 100   \\
			\rule{0pt}{10pt}
			$\theta$    &0.0001   &$\beta_{\mathrm{0}}$   &-30 dB    \\
			\rule{0pt}{10pt}
			$P$    &1000          &$\lambda$  &0.001   \\
			\rule{0pt}{10pt}
			$Number_e$    &1000   &$B$          & 100 MHz   \\
			\rule{0pt}{10pt}
			$\mu_1$    &0.99      &$\mu_2$    &0.95   \\
			\rule{0pt}{10pt}
			$P_f$    &0.11 W      &$P_h$      &0.08     \\
			\rule{0pt}{10pt}
			$F_{m,max}^{\rm MTU}$    &6 GHz      &$F_{j,max}$    &8 GHz     \\
			\rule{0pt}{10pt}
			$F_{max}^{\rm UAV}$  &10 GHz    &$V$   &20 m/s     \\	
			\hline
		\end{tabu}
	\end{center}
\end{table}

\begin{figure}[t]
	\includegraphics[width=3.6in,height=3.2in]{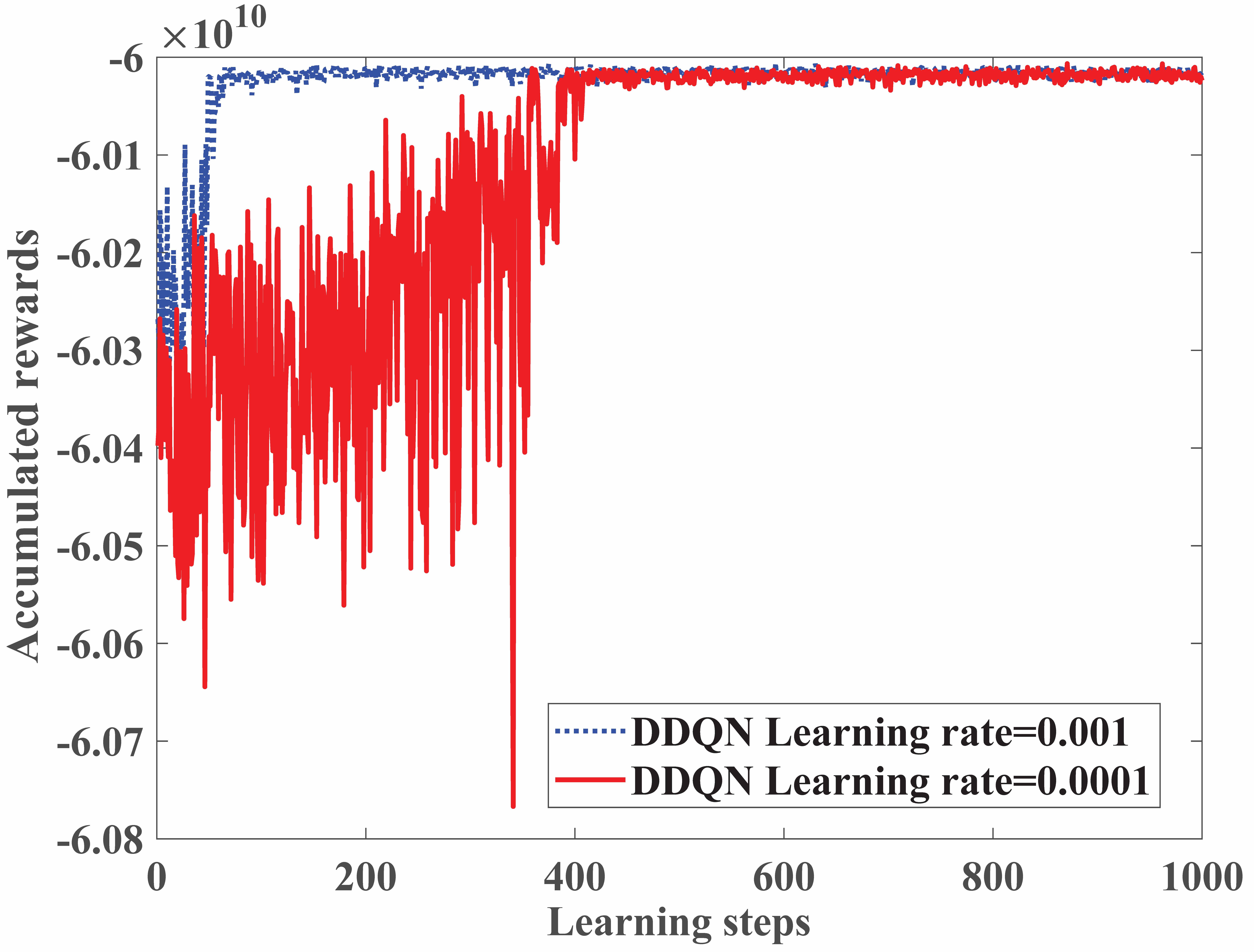}
	\setlength{\abovecaptionskip}{0.5pt}
	\caption{The convergence performance of DDQN with different learning rate $\lambda$.}
	\label{Compare}
\end{figure}

\begin{figure}[t]
	\includegraphics[width=3.6in,height=3in]{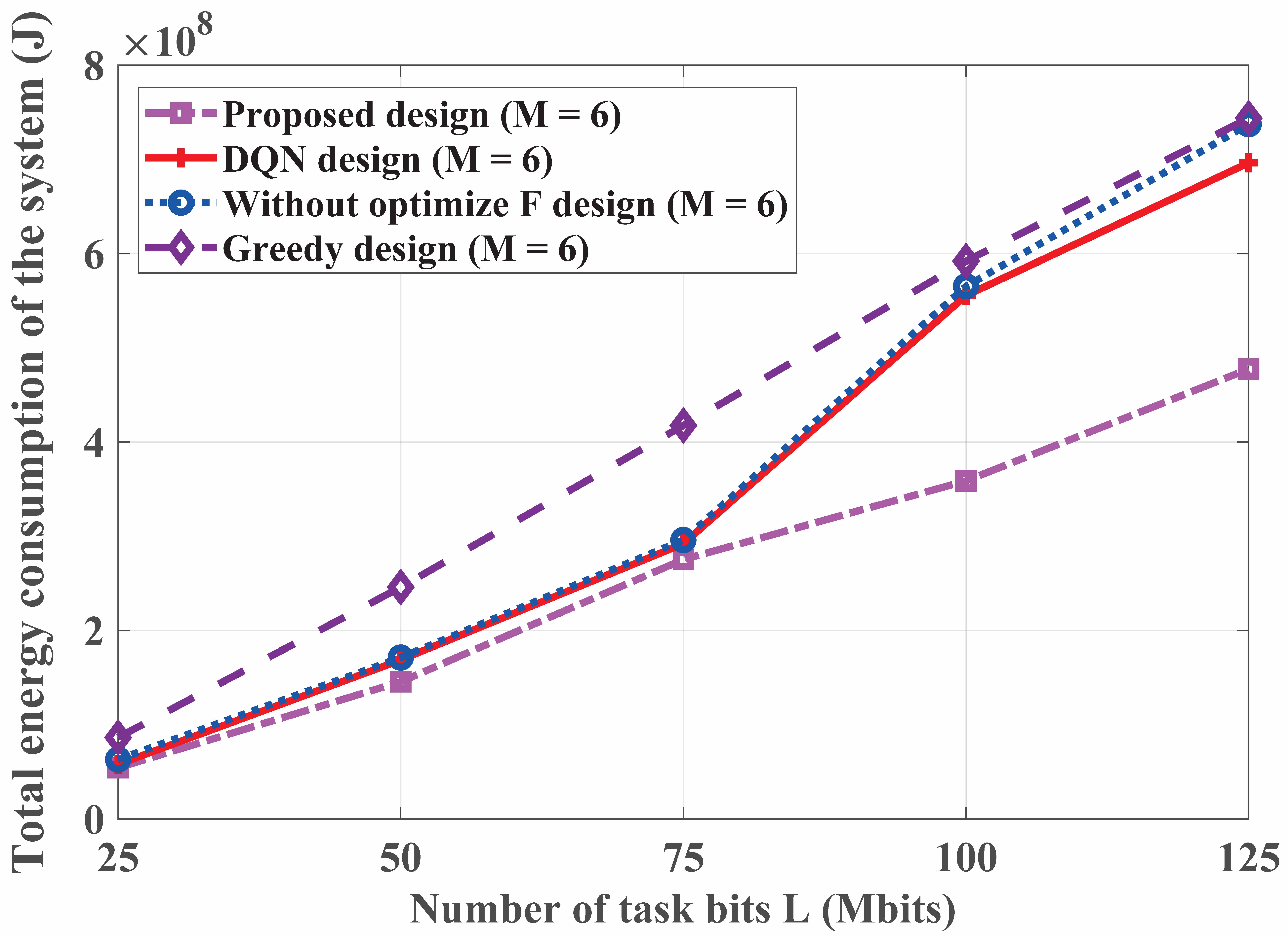}
	\setlength{\abovecaptionskip}{1pt}
	\caption{The energy consumption of the system vs. task quantity $L$.}
	\label{Task3}
\end{figure}

\begin{figure}[t]
	\setlength{\abovecaptionskip}{1pt}
	\includegraphics[width=3.6in]{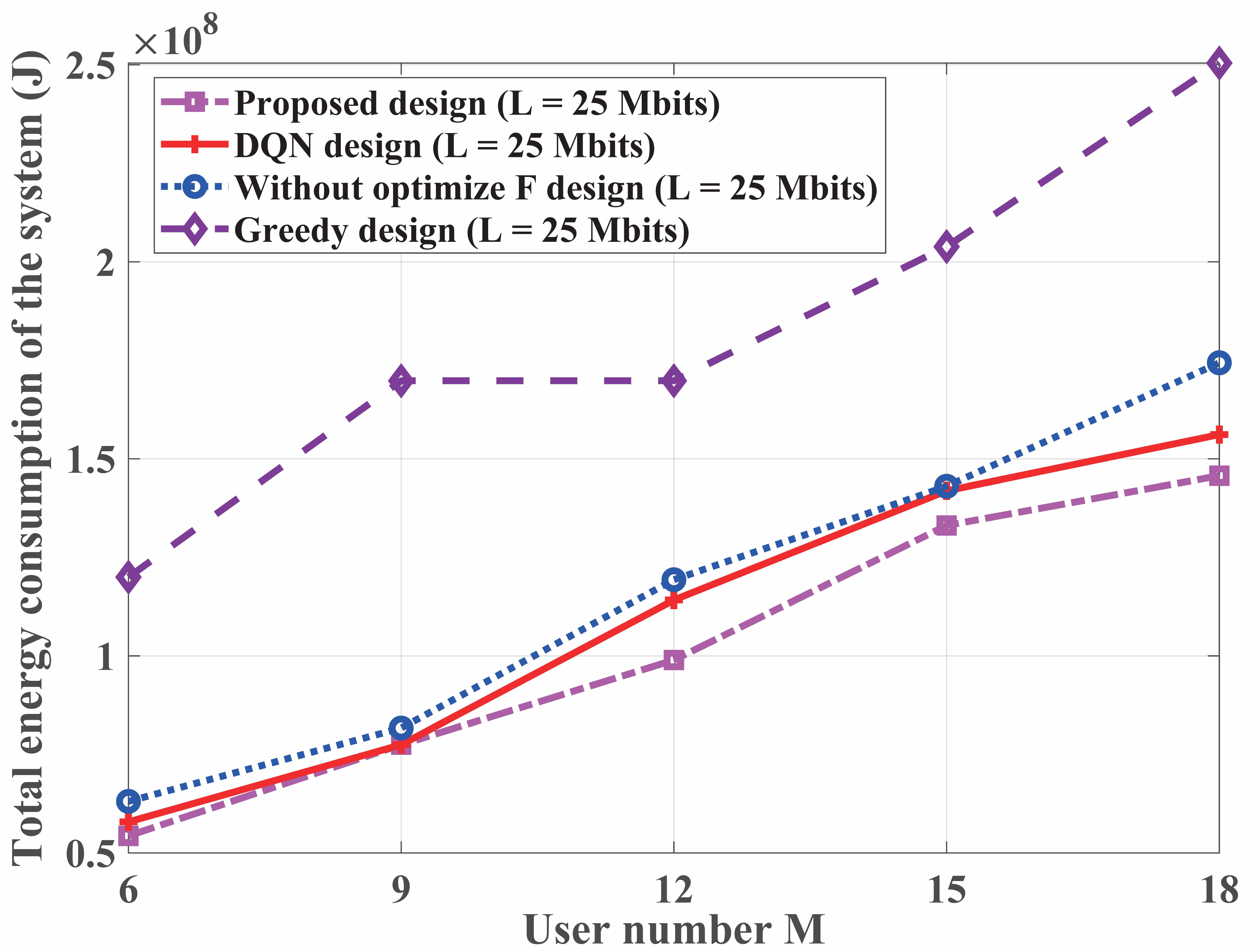}
	\caption{The energy consumption of the system vs. user number $M$.}
	\label{User3}
\end{figure}

\begin{figure}[t]
	\includegraphics[width=3.6in,height=3.05in]{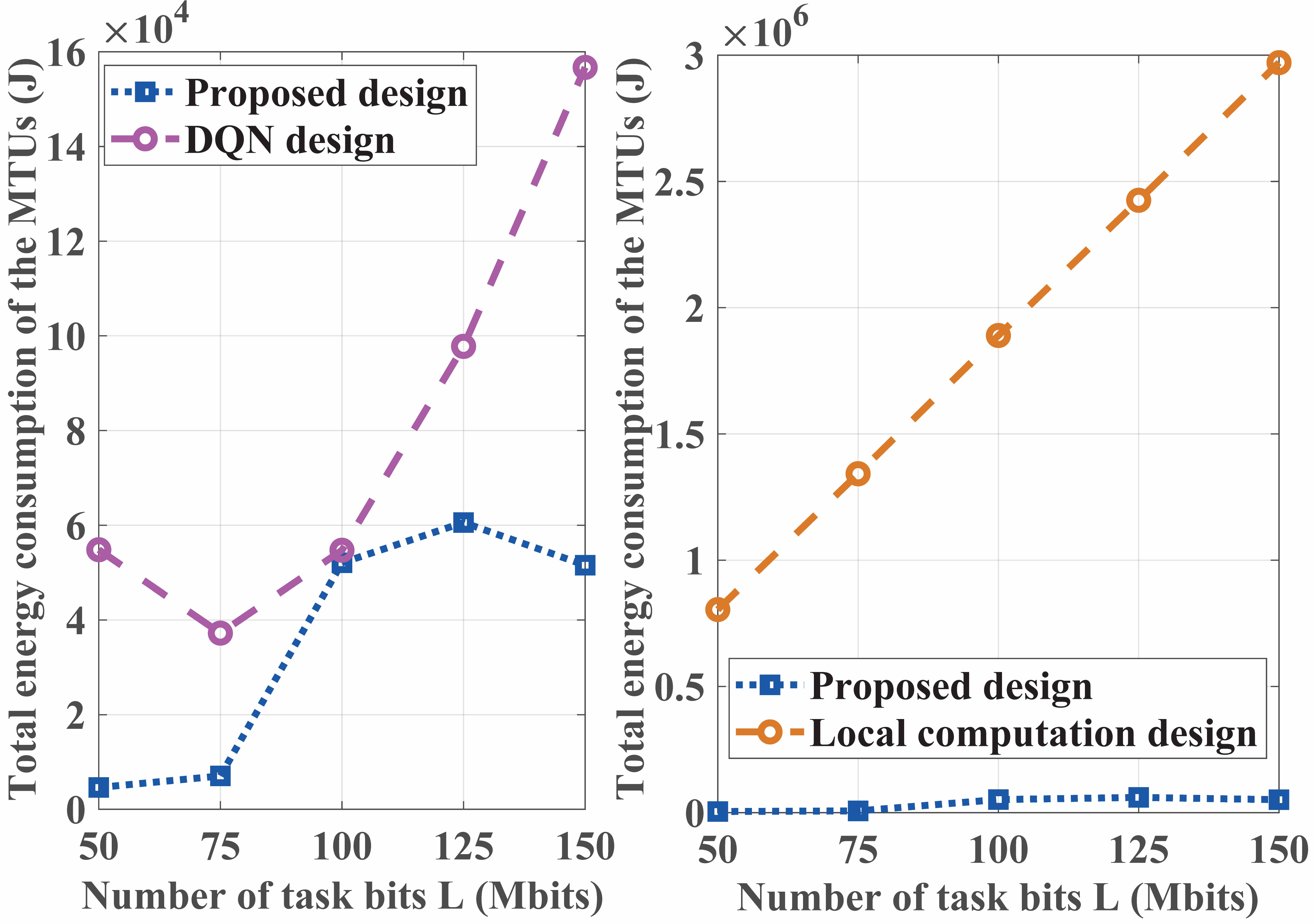}
	\setlength{\abovecaptionskip}{4pt}
	\caption{The energy consumption of MTUs vs. task quantity $L$.}
	\label{Local}
\end{figure}

\begin{figure}[t]
	\includegraphics[width=3.6in]{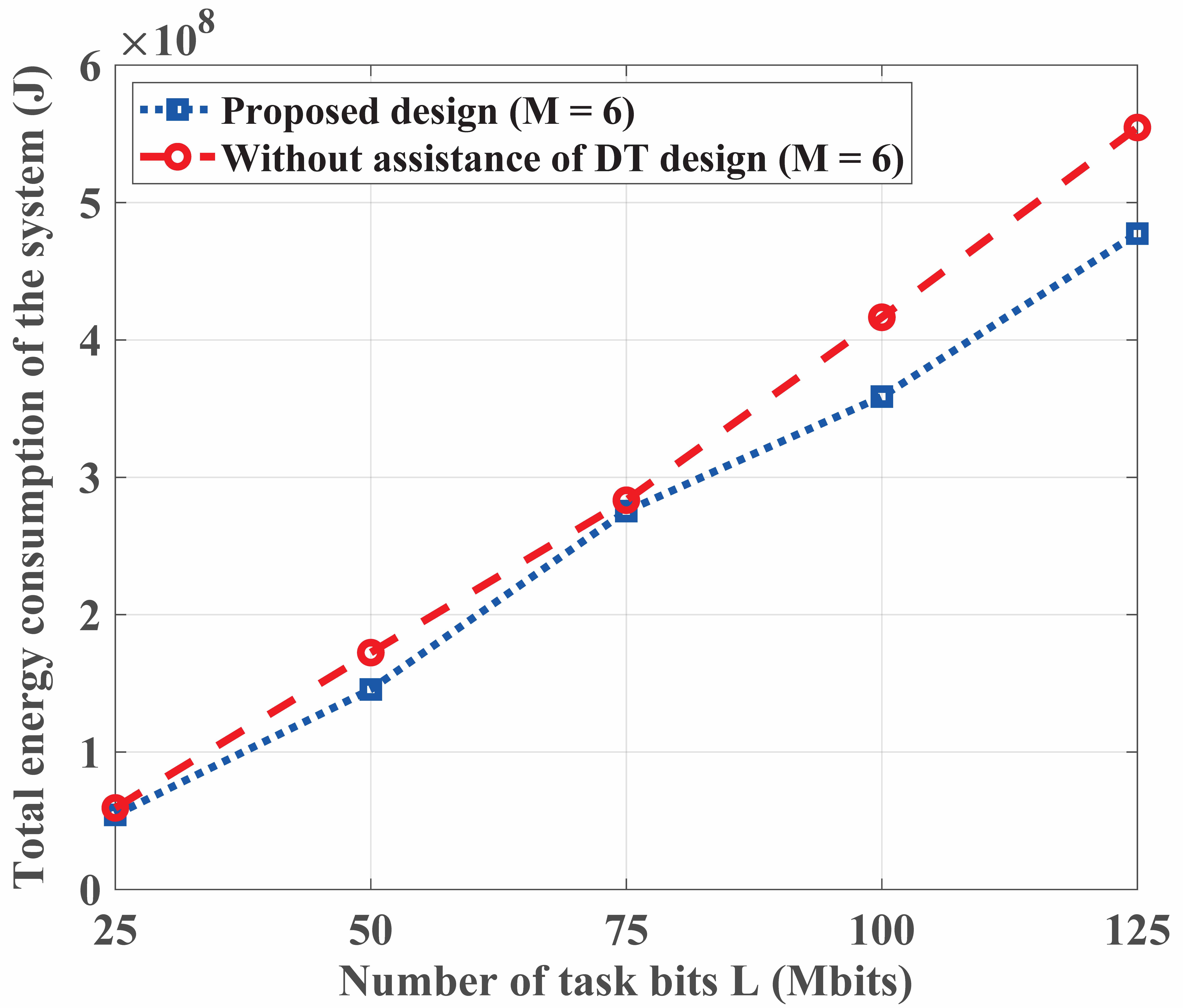}
	\setlength{\abovecaptionskip}{1pt}
	\caption{The energy consumption of the system vs. task quantity $L$.}
	\label{DT}
\end{figure}

\begin{figure}[h]
	\centering
	\includegraphics[width=3.6in]{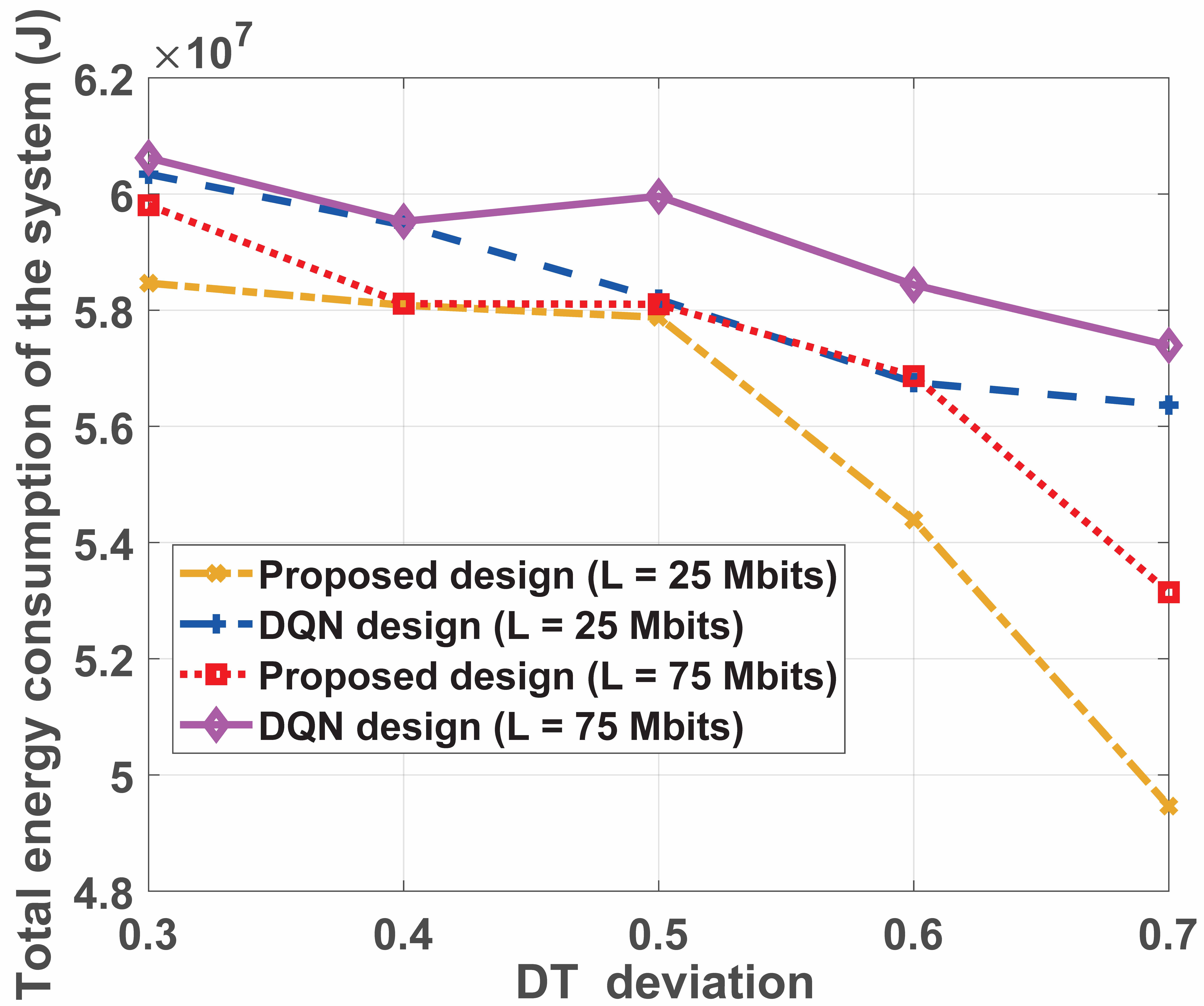}
	\setlength{\abovecaptionskip}{1pt}
	\caption{The energy consumption of the system vs. DT deviation.}
	\label{Compare_dev}
\end{figure}

\begin{figure}[t]
	\centering
	\includegraphics[width=3.6in]{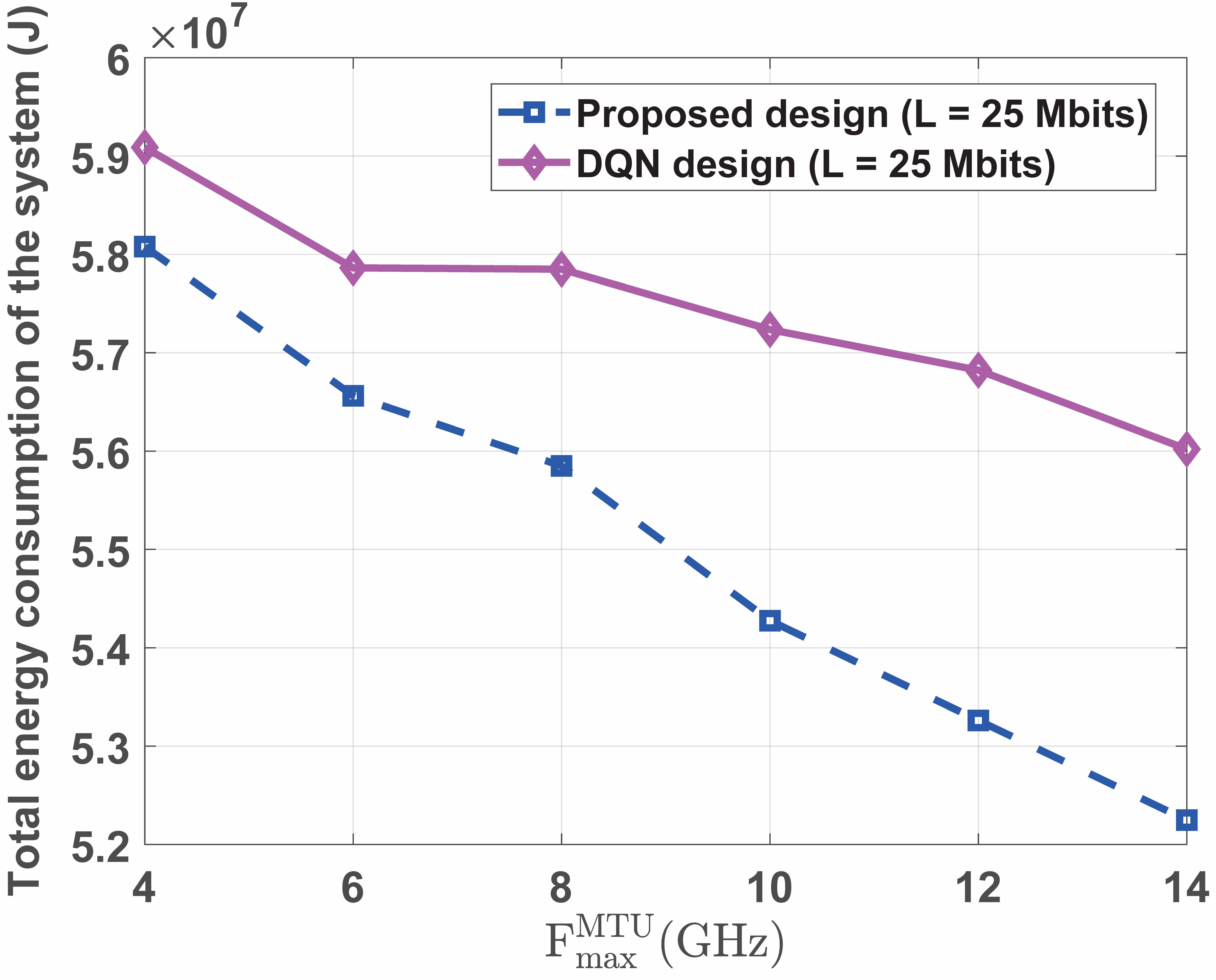}
	\setlength{\abovecaptionskip}{1pt}
	\caption{The energy consumption of the system vs. $\rm F_{max}^{MTU}$.}
	\label{Compare_f}
\end{figure}

Note that in order to illustrate the effectiveness of our proposed design, five benchmark schemes are designed as follows:

\begin{itemize}
	\item
	\textbf{DQN design:}
	In this case, the offloading decisions of all MTUs are optimized under DQN algorithm.
	
	\item
	\textbf{Without optimize F design:} In this case, all MTUs, UAV and resource devices computation capacity allocation F are not optimized.
	
	\item
	\textbf{Local computation design:}
	In this case, the tasks can only be computed locally at each MTU.

	\item
	\textbf{Greedy design:}
	In this case, the tasks of all MTUs can only be calculated at resource devices.
	
	\item
	\textbf{Without assistance of DT design:}
	In this case, the whole system is not assisted by DT.
\end{itemize}

Fig. \ref{Task3} depicts the total energy consumption of the system versus different required task quantity under $M = 6 $. 
As a whole, it is observed that the total energy consumption of the system increases with the increase of the number of task bits, regardless of the design scheme. Nevertheless, our proposed design always achieves the best performance compared with other designs, and the advantages become much more evident as the value of task quantity of each MTU increases. 
Compared with the design of DQN, it can be explained by the fact that DQN uses the same values to select and evaluate an action, but the proposed design overcomes the drawback and further improves the target value, so the proposed method can achieve better performance. 
Moreover, there is no doubt that the energy consumption of the system will be further increased if computation capacity allocation $\textbf{F}$ is not optimized.
In addition, it can easily notice that the energy costed by \textit{Greedy design} increases sharply with $L$ increasing. This is due to the fact that for some tasks, simply allowing them to be offloaded onto resource devices is not the best option, resulting in high energy consumption.

Fig. \ref{User3} compares the total energy consumption of the system under the proposed design and the benchmark scheme over different user number $M$.
From a vertical point of view, when the number of users remains unchanged, the total energy consumption of the system under the proposed design is always lower than that of the benchmark schemes, i.e., \textit{DQN design}, \textit{Without optimize F design} and \textit{Greedy design}.
Of all the designs, the \textit{Greedy design} achieves the highest energy consumption and increases dramatically as the number of users $M$ increases. Compared with the other three designs, resource device is the only choice for MTUs to complete tasks in this design, while the other three designs provide MTUs with a variety of choices. 
From a horizontal perspective, as the number of users $M$ increases, the overall energy consumption of the system continues to rise regardless of the design. The main reason is that the more MTUs there are, the more computing resources are required, which in turn increases the energy consumption of the whole process.
Also, the performance gap between our proposed design and other designs gets larger when the number of users increases.
Intuitively speaking, adding servers to assist MTUs in calculation can effectively achieve the goal of improving the performance of the system.
Therefore, our proposed design is more efficient when a large volume of MTUs exist in the system.

We further investigate the trend of the total energy consumption of the MTUs in Fig. \ref{Local}, where the number of task bits varies from $50$ Mbits to $150$ Mbits. It can be observed that the proposed design is superior to the \textit{Greedy design} and \textit{Local computation design}.
As expected, the other designs achieve more smaller value of energy consumption compared with the \textit{Local computation design}. This can be contributed to the fact that UAV, resource devices and BS as helpers can help task caculation. 
In addition, it can also be seen that compared with the \textit{Greedy design}, the proposed design achieves a further reduction in the total energy consumption of MTUs.

In Fig. \ref{DT}, we analyze the impact of the appearance of DT on the MEC system performance. As can be observed, the system performance measured by the energy consumption of system under DT assistance
is significantly better than that \textit{Without assistance of DT design}.
The reason for this phenomenon is that the states of each MTU, UAV and resource device are stored in DT, and no additional data interaction is required when search for offloading points. As a result, the energy consumption of the system is reduced and the time for transmitting data is saved. 

We then measure the energy consumption of the system over the varying DT deviation $\widetilde{f}_{m}^{l}$ and the task quantity $L$.
It can be clearly seen from Fig. \ref{Compare_dev} that our proposed design always achieves the best performance compared with DQN design. On one hand, when the task quantity $L$ is constant, the corresponding total energy consumption of the system is inversely correlated with the DT deviation. This can be explained from (14), (23) and (32) that a large positive error means the estimated value of DT is worse than the actual value, and thus the actual  energy consumption of the system is less than the estimated value. On the other hand, under the condition that the DT deviation is constant, as the task quantity $L$ increases, the corresponding system energy consumption will increase. This is because when the computing tasks increase, more data needs to be transmitted from MTUs to MEC serves due to the unsatisfied
latency constraints, which in turn consumes additional energy for data transmission.

Fig. \ref{Compare_f} compares the performance of the proposed scheme with the benchmark scheme over different maximal computation capacity at MTUs ($\rm F_{max}^{MTU}$). It can be observed that the total energy consumption of the system decreases with the increasing of $\rm F_{max}^{MTU}$. The main reason for this phenomenon is that when the maximal computation capacity at MTUs ($\rm F_{max}^{MTU}$) increases, MTUs can save more energy through local computation rather than offloading while guaranteeing their latency requirements.

\section{Conclusion \label{g}}
In this paper, we investigated a task offloading scheme in the DT-aided aerial edge computing and network, where MTUs randomly generate computing tasks in the process of moving and the high-performance edge nodes are exploited to help the task execution within a specified time.}   
To handle the design optimization problem, the DDQN algorithm was successfully applied to realize intelligent offloading of MTUs tasks and UAV deployment.
Furthermore, the closed-form expression was derived to quickly get the optimal transmission power distribution and an efficient iterative algorithm was used to achieve the computation capacity allocation of multiple MTUs, resource devices and UAV.  
Finally, numerical results were conducted to show that our proposed design can reduce the whole MEC system energy consumption by $7\%$, $11\%$ and $59\%$ compared with the DQN design, Without optimize F design and Greedy design, respectively. In future work, we will take into account the multi-device wireless interference model for a large number of devices.

\normalem
\bibliographystyle{IEEEtran}
\bibliography{refs}

\end{document}